\documentclass[ras]{agutex}

\def\fdg{\hbox{$.\!\!^\circ$}}
\def\farcm{\hbox{$.\!\!'$}}
\def\farcs{\hbox{$.\!\!''$}}
\usepackage{mathtools, amssymb, comment}

  \usepackage{graphicx, epstopdf}
  \setkeys{Gin}{draft=false}
\authorrunninghead{LOI ET AL.}

\titlerunninghead{MWA IONOSPHERIC POWER SPECTRUM ANALYSIS}

\authoraddr{Corresponding author: S.~T.~Loi,
Sydney Institute for Astronomy, School of Physics, The University of Sydney, NSW 2006, Australia. (sloi5113@uni.sydney.edu.au)}

\begin{document}

%
%

\title{Power spectrum analysis of ionospheric fluctuations with the Murchison Widefield Array}
%
%

%
%









\def\USydney{$^{1}$}
\def\CAASTRO{$^{2}$}
\def\Curtin{$^{3}$}
\def\CASS{$^{4}$}
\def\ANU{$^{5}$}
\def\ASTRON{$^{6}$}
\def\MIT{$^{7}$}
\def\UWisc{$^{8}$}
\def\SKASA{$^{9}$}
\def\CfA{$^{10}$}
\def\Rhodes{$^{11}$}
\def\ASU{$^{12}$}
\def\Haystack{$^{13}$}
\def\RRI{$^{14}$}
\def\UToronto{$^{15}$}
\def\UW{$^{16}$}
\def\Victoria{$^{17}$}
\def\UMichigan{$^{18}$}
\def\Tata{$^{19}$}
\def\NRAO{$^{20}$}
\def\UMelbourne{$^{21}$}

\author{
Shyeh Tjing Loi\USydney$^,$\CAASTRO,
Cathryn M.~Trott\CAASTRO$^,$\Curtin,
Tara Murphy\USydney$^,$\CAASTRO,
Iver H.~Cairns\USydney,
Martin Bell\CAASTRO$^,$\CASS,
Natasha Hurley-Walker\Curtin,
John Morgan\Curtin,
Emil Lenc\USydney$^,$\CAASTRO,
A.~R.~Offringa\CAASTRO$^,$\ANU$^,$\ASTRON,
L.~Feng\MIT,
P.~J.~Hancock\CAASTRO$^,$\Curtin,
D.~L.~Kaplan\UWisc, 
N.~Kudryavtseva\Curtin,
G.~Bernardi\SKASA$^,$\CfA$^,$\Rhodes,
J.~D.~Bowman\ASU, 
F.~Briggs\CAASTRO$^,$\ANU,
R.~J.~Cappallo\Haystack, 
B.~E.~Corey\Haystack, 
A.~A.~Deshpande\RRI, 
D.~Emrich\Curtin,
B.~M.~Gaensler\USydney$^,$\CAASTRO$^,$\UToronto, 
R.~Goeke\MIT,
L.~J.~Greenhill\CfA,
B.~J.~Hazelton\UW, 
M.~Johnston-Hollitt\Victoria,
J.~C.~Kasper\CfA$^,$\UMichigan, 
E.~Kratzenberg\Haystack, 
C.~J.~Lonsdale\Haystack, 
M.~J.~Lynch\Curtin, 
S.~R.~McWhirter\Haystack,
D.~A.~Mitchell\CAASTRO$^,$\CASS, 
M.~F.~Morales\UW, 
E.~Morgan\MIT, 
D.~Oberoi\Tata, 
S.~M.~Ord\CAASTRO$^,$\Curtin,
T.~Prabu\RRI, 
A.~E.~E.~Rogers\Haystack, 
A.~Roshi\NRAO, 
N.~Udaya~Shankar\RRI, 
K.~S.~Srivani\RRI, 
R.~Subrahmanyan\CAASTRO$^,$\RRI, 
S.~J.~Tingay\CAASTRO$^,$\Curtin, 
M.~Waterson\Curtin$^,$\ANU,
R.~B.~Wayth\CAASTRO$^,$\Curtin, 
R.~L.~Webster\CAASTRO$^,$\UMelbourne, 
A.~R.~Whitney\Haystack, 
A.~Williams\Curtin, 
C.~L.~Williams\MIT
}

$^{1}$Sydney Institute for Astronomy, School of Physics, The University of Sydney, NSW 2006, Australia\\
$^{2}$ARC Centre of Excellence for All-sky Astrophysics (CAASTRO)\\
$^{3}$International Centre for Radio Astronomy Research, Curtin University, Bentley, WA 6102, Australia\\
$^{4}$CSIRO Astronomy and Space Science (CASS), PO Box 76, Epping, NSW 1710, Australia\\
$^{5}$Research School of Astronomy and Astrophysics, Australian National University, Canberra, ACT 2611, Australia\\
$^{6}$Netherlands Institute for Radio Astronomy (ASTRON), Postbus 2, 7990 AA Dwingeloo, The Netherlands\\
$^{7}$Kavli Institute for Astrophysics and Space Research, Massachusetts Institute of Technology, Cambridge, MA 02139, USA\\
$^{8}$Department of Physics, University of Wisconsin--Milwaukee, Milwaukee, WI 53201, USA\\
$^{9}$Square Kilometre Array South Africa (SKA SA), Cape Town 7405, South Africa\\
$^{10}$Harvard-Smithsonian Center for Astrophysics, Cambridge, MA 02138, USA\\
$^{11}$Department of Physics and Electronics, Rhodes University, PO Box 94, Grahamstown, 6140, South Africa\\
$^{12}$School of Earth and Space Exploration, Arizona State University, Tempe, AZ 85287, USA\\
$^{13}$MIT Haystack Observatory, Westford, MA 01886, USA\\
$^{14}$Raman Research Institute, Bangalore 560080, India\\
$^{15}$Dunlap Institute for Astronomy and Astrophysics, The University of Toronto, ON M5S 3H4, Canada\\
$^{16}$Department of Physics, University of Washington, Seattle, WA 98195, USA\\
$^{17}$School of Chemical \& Physical Sciences, Victoria University of Wellington, Wellington 6140, New Zealand\\
$^{18}$Department of Atmospheric, Oceanic and Space Sciences, University of Michigan, Ann Arbor, MI 48109, USA\\
$^{19}$National Centre for Radio Astrophysics, Tata Institute for Fundamental Research, Pune 411007, India\\
$^{20}$National Radio Astronomy Observatory, Charlottesville and Greenbank, USA\\
$^{21}$School of Physics, The University of Melbourne, Parkville, VIC 3010, Australia\\ \vspace{-0.5cm}

%
%


\begin{abstract}
Low-frequency, wide field-of-view (FoV) radio telescopes such as the Murchison Widefield Array (MWA) enable the ionosphere to be sampled at high spatial completeness. We present the results of the first power spectrum analysis of ionospheric fluctuations in MWA data, where we examined the position offsets of radio sources appearing in two datasets. The refractive shifts in the positions of celestial sources are proportional to spatial gradients in the electron column density transverse to the line of sight. These can be used to probe plasma structures and waves in the ionosphere. The regional (10--100 km) scales probed by the MWA, determined by the size of its FoV and the spatial density of radio sources (typically thousands in a single FoV), complement the global (100--1000 km) scales of GPS studies and local (0.01--1 km) scales of radar scattering measurements. Our data exhibit a range of complex structures and waves. Some fluctuations have the characteristics of travelling ionospheric disturbances (TIDs), while others take the form of narrow, slowly-drifting bands aligned along the Earth's magnetic field.
\end{abstract}

%
%

%

\begin{article}

%
%

\section{Introduction}\label{sec:intro}

\subsection{Physical Background}
The ionosphere is the ionized component of the Earth's atmosphere. It lies between $\sim$50--1000\,km altitude, transitioning smoothly into the plasmasphere, which is a torus of fully-ionized plasma extending up to several Earth radii \citep{Singh2011}. At night, the electron density peaks near 300--400\,km \citep{Luhmann1995, Solomon2010}. The ionosphere and plasmasphere are embedded with complex density structures that can cause the refraction and diffraction of radio waves \citep{Darrouzet2009, Schunk2009}. These have been studied for many decades with low-frequency radio interferometer arrays \citep{Hewish1951, Bolton1953, Mercier1986, Jacobson1992, Jacobson1996, Helmboldt2012a, Helmboldt2014a}. 

Coupling between the various layers of the Earth's neutral atmosphere and ionosphere occurs through electric fields, ion-neutral collisions, shear flows, and upward-propagating acoustic-gravity waves (AGWs). These mechanisms, as well as the effects of thunderstorms, geomagnetic substorms, solar X-ray flares and large seismic events can produce irregularities in the ionosphere \citep{Rottger1978, Luhmann1995, Lastovicka2006, Liu2011}. Many travelling ionospheric disturbances (TIDs) are AGWs, which typically have wavelengths of 100--1000\,km and periods of minutes to hours \citep{Hocke1996}. These have been observed to produce quasi-periodic shifts in the positions of radio sources as the waves propagate across the field of view \citep{Bougeret1981}. 

Field-aligned density ducts are cylindrical enhancements/depletions in the plasmasphere with widths of 10--100\,km \citep{Angerami1970, Ohta1996}. These can extend down to ionospheric altitudes \citep{Bernhardt1977} and may form by the interchange of magnetic flux tubes under the influence of electric fields, e.g.~those from thunderstorms \citep{Park1971, Walker1978}. They have been reported to produce rapid variations in interferometer visibility phases as they drift relative to the background sources \citep{Jacobson1993, Hoogeveen1997a, Helmboldt2012b}.

\subsection{Technical Background}
The apparent angular offset in the positions of celestial radio sources (i.e.~refraction) from their true coordinates results from the distortion of radio-wave phase fronts due to differences in the electron column density (total electron content, TEC) between separate lines of sight to receivers \citep{Thompson2001, Schunk2009}. The angular offset is given by
\begin{equation}
  \Delta \theta \approx -\frac{1}{8\pi^2} \frac{e^2}{\epsilon_0 m_e} \frac{1}{\nu^2} \nabla_\perp \mathrm{TEC} \:, \label{eq:offset}
\end{equation}
where $e$ and $m_e$ are the electron charge and mass, $\epsilon_0$ is the vacuum permittivity, $\nu$ is the observing frequency, and $\nabla_\perp$TEC denotes the spatial gradient of the TEC transverse to the line of sight. The $\nu^{-2}$ dependence implies that offsets are larger at lower frequencies, and the negative sign indicates that sources refract towards lower TEC. The refractive wandering of a source over an integration due to temporal variations in the TEC can cause smearing and apparent decreases in peak flux density \citep{Kassim2007}. In addition, angular broadening and scintillation, which relate to the second spatial derivative of the TEC, occur when the angular sizes of density inhomogeneities are smaller than or comparable to the scattering angle \citep{Lee1975a, Lee1975, Rickett1977, Meyer-Vernet1980, Meyer-Vernet1981}. 

Methods for probing the ionosphere with radio waves include (1) ionospheric sounding, where the timing of broadband pulse reflections yields the vertical TEC profile \citep{Munro1950, Reinisch1986, Lei2005}; (2) radar scattering, where powerful narrowband pulses reveal small-scale TEC irregularities \citep{Gordon1958, Larsen2007}; (3) Global Positioning System (GPS) satellite measurements, where dual-frequency delays yield the absolute TEC along given lines of sight \citep{Mannucci1998, Lee2008}; (4) interferometer arrays, where TEC gradients are inferred from visibility phase fluctuations \citep{Hewish1951, Hamaker1978}; and (5) satellite-borne beacon receivers, which measure the TEC based on differential delays of signals from the ground and other satellites \citep{Bernhardt2006, Bernhardt2006a}. Radar scattering can detect structures 1--10\,m in size, over a region the width of the illuminating beam (several km). For GPS observations, the coverage and resolution depend on the distribution of ionospheric pierce points, which are the points where the lines of sight to ground stations intersect the ionosphere. Global GPS TEC maps with $\sim$100\,km resolution are publicly available, e.g.~from the Madrigal database \citep{Rideout2006}. Instruments aboard satellites in low-Earth orbit receiving signals from ground stations, such as the Scintillation and TEC Receiver in Space (CITRIS) instrument, have enabled the identification of scintillation-causing irregularities on scale sizes between 100\,m and 1\,km, and TEC measurements at a spatial resolution of 7.5\,km along the orbit path \citep{Siefring2011}. However, these do not currently provide as global a coverage as GPS.

Many radio interferometer studies of the ionosphere have involved tracking a single bright source and measuring the phase fluctuations on each baseline (where a baseline refers to a pair of interferometric receivers) as a function of time \citep{Hamaker1978, Jacobson1992a, Coker2009, Helmboldt2012a, Helmboldt2012}. The ionospheric phase $\phi$ ($\propto \Delta \theta$) is proportional to the difference in TEC between the lines of sight to the two receivers, so this can be used to probe temporal variations in the differential TEC on spatial scales corresponding to the projection of the array onto the ionosphere. We will refer to this approach as the \textit{antenna-based} method.

In antenna-based studies, the spatial sampling pattern is given by the array configuration, resulting in direction-dependent spatial resolution and a poor impulse response for power spectrum analysis of spatial features with sparse arrays \citep[e.g.][]{Helmboldt2014}. Widefield instruments that are able to observe many celestial radio sources simultaneously can overcome these limitations, because of the large number of isotropically-distributed pierce points. We will refer to the approach of analyzing source position shifts rather than antenna phases as the \textit{field-based} method. There have been several field-based studies of the ionosphere that used the Very Large Array (VLA) at 74\,MHz \citep{Kassim1993}. These include the work of \citet{Cohen2009}, who conducted a statistical analysis of pairwise differential offsets (3--8 sources in a FoV), and \citet{Helmboldt2012b}, who performed a spectral analysis of the position offsets of 29 bright sources in a single FoV and detected groups of wavelike disturbances. A combination of antenna- and field-based techniques has been used to conduct a climatological study of ionospheric irregularities using the VLA, characterizing structure on scales of several to hundreds of kilometers \citep{Helmboldt2012c}.

The Murchison Widefield Array \citep[MWA;][]{Lonsdale2009, Bowman2013, Tingay2013} is a 128-element, low-frequency (80--300\,MHz) dipole array located at the Murchison Radio-astronomy Observatory (MRO) in Western Australia. Each receiver element, called a ``tile'', comprises 16 dual-polarization bowtie dipole antennas arranged in a $4 \times 4$ configuration. The 128 tiles are spread over a 3 km-wide area in a pseudo-random but centrally condensed configuration, yielding a very dense instantaneous $u,v$-coverage which gives it an excellent snapshot imaging capability. It has a $\sim$30$^\circ$-wide FoV, enabling it to image a patch of the ionosphere about 150\,km across (assuming an ionospheric altitude of 300\,km). Over 1000 unresolved point sources are typically visible in a single 2-min integration, yielding pierce-point separations of 1--10\,km. With a high imaging cadence of up to 0.5\,s (the cadence at which the raw visibilities are stored, which sets the minimum integration time of each snapshot), the MWA enables us to study ionospheric structures and dynamics on regional scales at high spatiotemporal resolution. This complements the global scales and lower spatial resolutions of GPS measurements, and the highly localized probing of radar scattering experiments. Airglow imaging \citep{Makela2006, Shiokawa2013} can achieve widefield coverage similar to the MWA, but its reliance on optical wavelengths restricts operations to cloudless and moonless nights. Radio telescopes suffer no such limitations; radio observations can even be conducted in the daytime. Furthermore, airglow measurements are restricted to lower altitudes, whereas radio interferometers can detect irregularities over large altitude ranges up to high altitudes (in the topside region and plasmasphere) which are otherwise difficult to access from the ground \citep{Jacobson1996, Loi2015}.

The advent of sensitive, low-frequency, widefield radio instruments such as the MWA, the Low Frequency Array \citep[LOFAR;][]{vanHaarlem2013} and the Long Wavelength Array \citep[LWA;][]{Ellingson2009} gives us an opportunity to study the ionosphere in unprecedented spatial detail and breadth. Characterization of the ionosphere is relevant not only to ionospheric science but also to radio astronomy, because fluctuations in position and flux density can complicate the cross-matching and classification of radio sources. The ionosphere can also cause time- and position-dependent distortions of the telescope point-spread function (PSF), demanding non-trivial calibration and deconvolution schemes \citep{Mitchell2008, Intema2009, Bernardi2011}. Information about the expected spatial and temporal scales of these fluctuations is important for designing and optimizing algorithms that identify and remove ionospheric distortions. 

\subsection{This Work}
We present the results of a power spectrum analysis of ionospheric fluctuations observed using the MWA. We analyzed the position offsets of point sources appearing in two MWA datasets from the commensal Epoch of Reionization (EoR) Transients Survey at 183\,MHz (G0005; PIs: P.~J.~Hancock, L.~Feng \& N.~Kudryavtseva) and the MWA Transients Survey (MWATS) at 154~MHz (G0001; PI: M.~Bell), hereafter datasets A and B respectively, as a function of sky position and time. 

This paper is structured as follows. In \S\ref{sec:method}, we describe the data reduction process and our approach to computing the power spectra. In \S\ref{sec:results}, we demonstrate this on MWA data from the two aforementioned datasets. We interpret the observed features in \S\ref{sec:discussion}. Finally, we conclude in \S\ref{sec:conclusions}.

\section{Method}\label{sec:method}
\begin{figure}[H]
  \centering
  \includegraphics[width=0.5\textwidth]{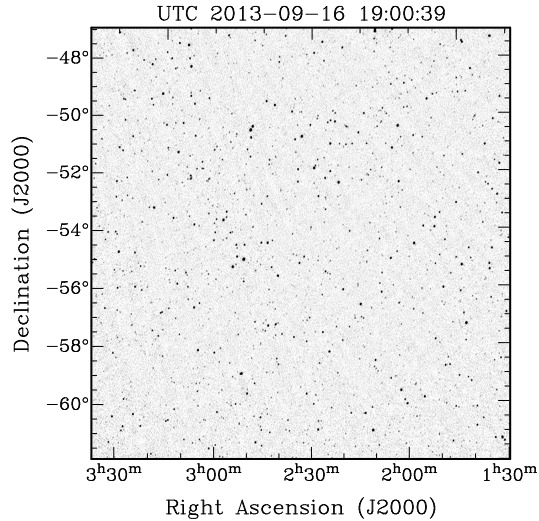}
  \caption{Example snapshot image taken by the MWA (dataset B, epoch 53), zoomed in to the central one-third so that individual sources are visible. Each of the $\sim10^3$ celestial radio sources in each snapshot image contributes a pierce point. The intensity is displayed on a linear greyscale.}
  \label{fig:snapshot}
\end{figure}

All MWA data mentioned in this work were obtained at night. This work uses radio images synthesized using the entire array (field-based method), rather than information from individual baselines (antenna-based method). The whole MWA can be considered as a single receiver, and the number of pierce points is equal to the number of celestial radio sources visible at any given epoch (e.g.~Figure \ref{fig:snapshot}). The antenna-based method, which needs to observe a single bright source that dominates the visibilities, is unsuitable for the MWA because of its large FoV and thereby large number of bright sources in any given field. Since the position offset of a source in an image formed using all 128 tiles is related to the average ionospheric phase over all baselines, the field-based approach effectively smooths structure in the ionosphere by a window of diameter equal to the physical dimensions of the MWA. The fairly compact size of the array (longest baseline $\sim$3\,km) means that our angular resolution for ionospheric work is about 0\fdg6, comparable to the average angular separation of point sources at typical instrument sensitivities. This is about an order of magnitude lower than the angular resolution of the images, which is set by the size of the MWA synthesized beam ($\sim$2--3\,arcmin).

\subsection{MWA Data Reduction}\label{sec:datareduction}
The process of synthesizing an image from interferometric visibilities involves a Fourier inversion of the spatial coherence function of the electric field, which is sampled discretely by the individual baselines \citep{Thompson2001}. Prior to the inversion step it is often necessary to excise (``flag'') unusable portions of the data, for example those contaminated by radio-frequency interference (RFI), and also to calibrate for varying instrumental gains, which might arise for example from thermally-induced changes in cable lengths. Several algorithms exist for performing instrument gain calibration. These include phase referencing \citep{Fomalont1999}, which involves observing a well-understood source of emission (a calibrator) and solving for the antenna gains that minimize the overall differences between measured and expected values, and self-calibration \citep{Pearson1984, Cornwell1999}, which exploits a set of closure relations that relate the gains of each closed loop of antennas in an over-determined fashion to solve for individual antenna gains.

Fourier transforming the gridded and weighted visibilities produces an image (the ``dirty'' image) that is the true sky brightness distribution convolved with the telescope PSF. The shape of the PSF is determined by the sampling pattern of the spatial coherence function (the $u,v$-coverage) and in general has a complicated shape. Various deconvolution algorithms exist for recovering the true brightness distribution from a dirty image. Point-like (unresolved) sources, such as those considered for analysis here, can be deconvolved effectively using the \textsc{clean} algorithm \citep{Hogbom1974}. This involves the iterative subtraction of a scaled version of the PSF from local intensity maxima, until certain convergence criteria are satisfied, to obtain a set of \textsc{clean} components representing the point sources. These components are reconvolved with a two-dimensional Gaussian whose width is representative of the angular resolution of the telescope and added back to the residuals of the deconvolution step, yielding an idealized approximation to the sky brightness distribution (the ``restored'' image).

The two datasets discussed in this work were originally obtained by the MWA for different scientific purposes (neither ionospheric), under different observing strategies and reduced by separate science teams using different pipelines. We have appropriated these data for the current work; these differences are not intentional and are beyond the scope of control of our study.

Dataset A was taken on 2014-02-05 at 183\,MHz with 30.72\,MHz bandwidth, in ``drift-and-shift'' mode, in which the MWA analog beamformers point towards a new azimuth and elevation approximately every half hour, and a consistent patch of sky then drifts through the FoV. This minimizes the number of pointings used, while still tracking the same patch of sky. Two-minute snapshots were continuously taken of the same patch of sky over the observing duration, which was centred near local midnight (see Table \ref{tab:datasets} for details). 

Known failed antennas were flagged, comprising 12\% of the baselines. The data were flagged for RFI using the automated flagging routine \textsc{aoflagger} \citep{Offringa2012}, removing 1\% of the visibilities. They were time-averaged from 0.5\,s to 4\,s and were not smoothed from their native 40\,kHz frequency resolution. The data were calibrated from an observation of 3C444 (RA=$22^{\mathrm{h}}14^{\mathrm{m}}25.752^{\mathrm{s}}$; Dec=$-17^{\circ}01'36\farcs29$), assuming a point-source model. This has a flux density of $67$\,Jy and a spectral index of $-1.0$ at 183\,MHz \citep{Hurley-Walker2014}, sufficient to dominate the visibilities and produce a set of per-channel, per-polarization, per-antenna complex gains.

After applying these gains, the data were imaged using multi-frequency synthesis across all 30.72\,MHz of bandwidth using \textsc{wsclean} \citep{Offringa2014}, down to the first negative \textsc{clean} component. This initial model was Fourier-transformed to produce a model visibility set. These were then applied in one round of self-calibration to obtain a single solution for the amplitude and phase of each antenna over the whole observing interval. This correction is position- and time-independent: it should have no effect on the relative positions of the field sources, and should only act to reduce instrumental errors. Note that unlike for conventional instruments with much narrower fields of view, solely antenna-based corrections are ineffective at removing ionospheric phases from MWA data, because its wide FoV implies that the ionosphere cannot be treated as locally uniform within each FoV (and therefore that antenna and ionospheric phases are non-degenerate). The compact size of the MWA ensures that the effective antenna-based ionospheric phasor, which is a visibility-weighted average of celestial source ionospheric phasors over the FoV, is almost identical for each antenna thus allowing instrumental errors to dominate antenna-based phase differences. 

The data were then imaged in four 7.68\,MHz sub-bands. The pixel resolution in all imaging steps was 0\farcm75 and imaging included the entire FoV down to the first null of the primary beam (full width at first null $\sim$50$^\circ$). This yielded images of size $5000 \times 5000$ pixels. The images were produced in instrumental Stokes (the E-W dipole response, N-S dipole response, and the cross-terms). Using a model of the MWA primary beam \citep{Sutinjo2014}, these instrumental polarizations were then combined to produce Stokes~I images for onward analysis.

Dataset B was obtained on 2013-09-16 at 154\,MHz over a 30.72\,MHz bandwidth using drift-scan observations at a declination of $-55\fdg0$ on the meridian. A two-minute snapshot was obtained every six minutes at this declination for an entire night of observations (see Table 1 for details). These data were flagged for RFI (using \textsc{aoflagger}) and known problematic tiles. They were then averaged by a factor of two in time and four in frequency to yield a time resolution of 1\,s and a frequency resolution of 160\,kHz.

Frequency-dependent, time-independent calibration solutions as a function of tile and polarization were derived from an observation of PKS 2356$-$61 using a source model extrapolated from 843\,MHz \citep{Mauch2003}, and applied to all target observations. This has a peak flux density of 30.6\,Jy (integrated flux density of 114.6\,Jy) at 154\,MHz and a spectral index of $-0.85$. No self-calibration was performed on this dataset. The full-bandwidth data (30.72\,MHz) were then imaged and deconvolved using the \textsc{wsclean} algorithm \citep{Offringa2014} to yield images of size $3072 \times 3072$ pixels, the pixel size being $0\farcm75$. The data were \textsc{clean}ed up to a maximum of 20,000 iterations. The full 30.72\,MHz bandwidth was imaged. For each observation, images for two instrumental polarizations (the N-S and E-W dipole responses) were corrected for the primary beam and then combined \citep[for details, see][]{Offringa2014} to produce a Stokes I image.

\subsection{Source Extraction and Selection}\label{sec:sourceextraction}
We chose dataset A for discussion in this work because the sources within it exhibited a moderately large root-mean-square (RMS) amplitude of position shifts, this being 19\,arcsec on average and high compared to other datasets (characteristic RMS amplitude of 10--20\,arcsec at similar frequencies). This is smaller than the pixel size (45\,arcsec) and the synthesized beam FWHM (119\,arcsec at 183\,MHz). We used the source finder \textsc{Aegean} \citep{Hancock2012} to extract sources from the images, obtaining 1300--2100 sources per snapshot for a clip level of 5$\sigma$. Candidate sources were identified by \textsc{Aegean} as local regions of negative spatial curvature whose maxima lay above a given noise threshold. Their parameters were then computed by fitting two-dimensional Gaussians to the flux density distribution at those locations. An estimate for the position error associated with fitting a Gaussian to an unresolved source \citep{Condon1997} is
\begin{equation}
  \Delta \psi \approx \frac{\theta_b}{\text{SNR} \sqrt{8 \ln 2}} \:, \label{eq:poserror}
\end{equation}
where $\theta_b$ is the FWHM of the synthesized beam, assumed to be circular, and SNR refers to the signal-to-noise ratio, which we take to be the ratio of the fitted peak flux density to the local RMS noise. Note that positions can be measured to significantly greater precision than the instrument resolution, and that this precision increases with the SNR of the source (e.g.~for a source at the 5-$\sigma$ detection threshold, its position can be measured correct to $\sim$10\,arcsec, an order of magnitude smaller than the synthesized beamwidth). We excluded candidate sources with negative fitted peak flux densities, and those with large position errors ($>$1 arcmin) as quoted by \textsc{Aegean}. When \textsc{Aegean} encounters an island of pixels for which there are fewer pixels than fit parameters (6 parameters per component), the source parameters are estimated rather than fitted. We excluded such sources from our analysis. 

Cross-matching the remainder with the NRAO VLA Sky Survey (NVSS) catalog \citep{Condon1998} using a cross-matching radius of 1.2\,arcmin resulted in the identification of 4713 distinct sources and 55070 occurrences of any source in any epoch, implying that each source was observed in around 12 epochs on average. This dataset contains 38 epochs in total. Cross-matching was performed by identifying the nearest NVSS source within the threshold radius of a given \textsc{Aegean} source and associating it to the latter. Sources extracted by \textsc{Aegean} for which there was no NVSS source within the search radius were discarded. Around 7\% of sources were excluded in this manner. The purpose of cross-matching with an external catalog was to reduce the fraction of spurious sources, which may arise from noise spikes or residual sidelobes.

We chose dataset B to contrast dataset A in that the average amplitude of position offsets was low (9\,arcsec) compared to what is typically measured in other datasets, about half the amplitude seen in dataset A. Source extraction returned 1600--2800 sources per snapshot for a clip level of 5$\sigma$. We subjected the sources to the same filtering and cross-matching criteria as for dataset A, except that we used the Sydney University Molonglo Sky Survey (SUMSS) catalog \citep{Mauch2003} for cross-matching to suit the relatively southern declination. A total of 19979 distinct sources were identified, and there were 179450 occurrences over all epochs. Each source was thus observed in around 9 epochs on average. There are 77 epochs in total for this dataset. Sources for which there was no SUMSS match were discarded, this excluding $\sim$1\% of sources. 

Differences between the two datasets in the number of sources extracted per snapshot may be attributed to differing primary beam sizes (larger at lower frequencies), and differences in imaging and calibration approaches yielding different sensitivities. Differences in the cross-matching efficiency between the two datasets may be due to differing completeness levels of the reference catalogs, or perhaps higher levels of ionospheric activity in dataset A causing sidelobe degradation and an increase in spurious source identifications compared to dataset B.

\subsection{Analysis of Position Offsets}\label{sec:scalarfield}
The power spectrum analysis technique described here is broadly applicable to any scalar quantity measurable over the pierce points. In this work we analyzed just position offset information, where we considered scalar projections of the position offset vector field onto different directions on the sky (we obtained these by taking the inner product of each vector with a unit vector pointing in a chosen direction). We computed the position offset vectors as the displacements of sources from their time-averaged celestial coordinates, which we took to be the reference points. The technique can also be applied to source flux densities, polarization vectors or rotation measures, but we defer these to future work.

The purpose of using the time-averaged position of each source as a reference was to remove calibration-related position errors. We checked for these in separate analysis by cross-matching the sources in each dataset with the International Celestial Reference Frame Second Realization (ICRF2) catalogue \citep{Fey2009}. Many datasets displayed collective shifts of all sources from their ICRF2 positions, in some cases by up to one arcminute, where the direction of the shift depended on the calibrator. Since the calibration solutions have no time dependence, measuring the offsets with respect to the time-averaged position removes systematic position offsets arising from calibration errors, which are likely to produce spurious peaks in the power spectra. However, the disadvantage of this operation is that ionospheric fluctuations whose drift speeds happen to match those of the sources are subtracted away from the data. The use of a time average also alters the temporal spectral response, since this is akin to a detrending/high-pass filtering operation with an effective window size given by the duration over which a source appears ($\sim$12\,epochs = 24\,min on average for dataset A, and $\sim$9\,epochs = 54\,min for dataset B). The temporal frequency resolutions for the two datasets therefore differ from one another (roughly twice as high in dataset B than A), and are systematically lower than given by the reciprocals of the total observing durations. However, since many sources are detected for longer durations than average, the filtering effect is expected only to be partial and some sensitivity to lower temporal frequencies would be retained.

We specified pierce-point locations in terms of a two-dimensional Cartesian coordinate system whose $x$-axis points west and $y$-axis points north. These were obtained via an equiareal projection of the celestial coordinates of each radio source. The native spherical coordinates for the projection \citep[cf.][]{Calabretta2002}, whose longitude and latitude we refer to as $\beta$ and $\gamma$ respectively, are defined such that the fiducial point $(\beta_0, \gamma_0) = (0,0)$ corresponds to the zenith and the South Celestial Pole lies at $(\beta_p, \gamma_p) = (90^\circ-\Lambda, 0^\circ)$. Here $\Lambda$ denotes the latitude of the observing site, measured positive southwards; for the MWA, $\Lambda = 26\fdg7$. We have used the symbols $(\beta,\gamma)$ rather than the usual $(\phi,\theta)$ of \citet{Calabretta2002} to avoid clashes with other variables defined in this paper. Geometrically, this corresponds to a 90$^\circ$ rotation of the Az/El coordinate system so that the coordinate equator lies on the meridian.

The coordinates $(x,y)$ were computed using the pseudocylindrical Sanson-Flamsteed projection
\begin{equation}
    \begin{pmatrix}
    x \\
    y
  \end{pmatrix} = \begin{pmatrix}
    -\gamma \\
    (\beta_1 - \beta) \cos \gamma - \beta_1
  \end{pmatrix} \:, \label{eq:SFLproj}
\end{equation}
where $(\beta,\gamma) = (\beta_1,0)$ is the center of projection; here we chose $\beta_1$ to be the median value of $\beta$ over all pierce points in each dataset. With this choice of coordinates the $x$-axis points west, the $y$-axis points north, and the origin $(x,y) = (0,0)$ lies at the zenith. The Sanson-Flamsteed projection preserves spatial scales, although it fails to preserve angles. Angular distortions are minimized by our choice of native coordinates, whose poles are shifted away from the meridian where most MWA observations are conducted. A desirable property of these coordinates is that they are fixed to the terrestrial rather than the celestial sky, meaning that the velocity of a waveform will be measured with respect to the ground rather than the pierce points (which drift at $\sim$20\,m\,s$^{-1}$ through the ionosphere at an altitude of 300\,km).

For a given point source whose position offset vector at some epoch is $(\Delta x, \Delta y)$, we defined its projection onto the $\theta$-direction by
\begin{equation}
  s_\theta = \Delta x \cos \theta + \Delta y \sin \theta \:, \label{eq:projection}
\end{equation}
where the angle $\theta$ is measured anticlockwise from the positive $x$-axis ($\theta$ can be thought of as the polar angle in 2D cylindrical coordinates). The scalar $s_\theta$ is the quantity analyzed in all subsequent sections. 

\subsection{Power Spectrum Computation}\label{sec:computation}
Every occurrence of a known NVSS or SUMSS source at a given epoch in a dataset induces a corresponding pierce point defined in terms of one temporal ($t$) and two spatial $(x, y)$ coordinates. The set of pierce points $Q$, along with their associated $s_\theta$ values computed as per Equation (\ref{eq:projection}) for a chosen value of $\theta$, were then used to populate a three-dimensional, rectilinear, evenly-spaced datacube $M$ according to 
\begin{equation}
  M_{ijk}(\theta) = \begin{cases}
    s_\theta(q) & q = (x,y,t) \in Q \\
    0 & \text{otherwise.}
  \end{cases} \label{eq:datacube}
\end{equation}
Here $i, j, k \in \mathbb{Z}$ are the array indices of $M$ along the $x$, $y$ and $t$ directions, respectively. These were obtained from the spacetime coordinates of the pierce point $q = (x, y, t)$ by gridding according to
\begin{eqnarray}
  i(q) &=& \big\lceil (x - x_\text{min})/d \big\rceil \nonumber \\
  j(q) &=& \big\lceil (y - y_\text{min})/d \big\rceil \nonumber \\
  k(q) &=& t \:, \label{eq:coord2ind}
\end{eqnarray}
where $d$ is the chosen grid resolution, $x_\text{min}$ and $y_\text{min}$ are the minimum $x$ and $y$ values over $Q$, and $t$ is the epoch number. Note that it is impossible to ensure that the mapping in (\ref{eq:coord2ind}) is unique. Where clashes occurred, we chose to average the $s_\theta$ values together (this occurred at a low rate of 2--3\%).

The spatiotemporal power spectrum $\mathcal{P}$ for each dataset was then calculated by taking the squared modulus of the discrete Fourier transform of $M$ with respect to each of its three axes. We normalized the power spectrum by the observing volume $V$ (size of the Fourier datacube, with units of deg$^{-2}$ s$^{-1}$) and the number of pierce points $|Q|$ to obtain the dimensional power spectral density (units of deg$^4$ s). This is
\begin{eqnarray}
  \mathcal{P}_{\ell mn}(\theta) &=& \frac{1}{V |Q|} \Bigg| \sum_{i=1}^{N_x} \sum_{j=1}^{N_y} \sum_{k=1}^{N_t} M_{ijk}(\theta) \times \nonumber \\
   & & \exp\left[ -2\pi\iota \left( \frac{(i-1)(\ell-1)}{N_x} + \frac{(j-1)(m-1)}{N_y} + \right. \right. \nonumber \\
   & & \left. \left. \frac{(k-1)(n-1)}{N_t} \right) \right] \Bigg|^2 \:, \label{eq:powerspec}
\end{eqnarray}
where $N_x$, $N_y$ and $N_t$ are the dimensions of $M$ along the $x$, $y$ and $t$ axes, respectively, $\ell, m, n \in \mathbb{Z}^+$ index the $k_x$, $k_y$ and $\omega$ dimensions of the output datacube $\mathcal{P}$, and we have denoted $\iota \equiv \sqrt{-1}$ to avoid clashes with the symbols used for the indices of $M$. Here $\omega$ is the temporal frequency, $k_x$ is the E-W spatial frequency, and $k_y$ is the N-S spatial frequency. 

The temporal grid resolution for each dataset in dimensional units is simply equal to the snapshot cadence, which is 2\,min for dataset A and 6 min for dataset B. To choose a spatial grid resolution, we considered the angular size of the MWA projected onto the ionosphere, since this represents the size of the window over which differential baseline phases are averaged. Given a longest baseline of about 3\,km and assuming an altitude of 300\,km, this corresponds to a resolution of 36\,arcmin and a Nyquist sampling limit of 18\,arcmin. We chose a spatial increment of $d = 9$\,arcmin to grid the pierce points, which corresponds to spatially oversampling by a factor of two. The sizes of the resulting datacubes were $N_x \times N_y \times N_t = 264 \times 229 \times 38$ elements for dataset A and $262 \times 262 \times 77$ elements for dataset B.

\section{Results}\label{sec:results}
\begin{figure}[H]
  \centering
  \includegraphics[width=0.49\textwidth]{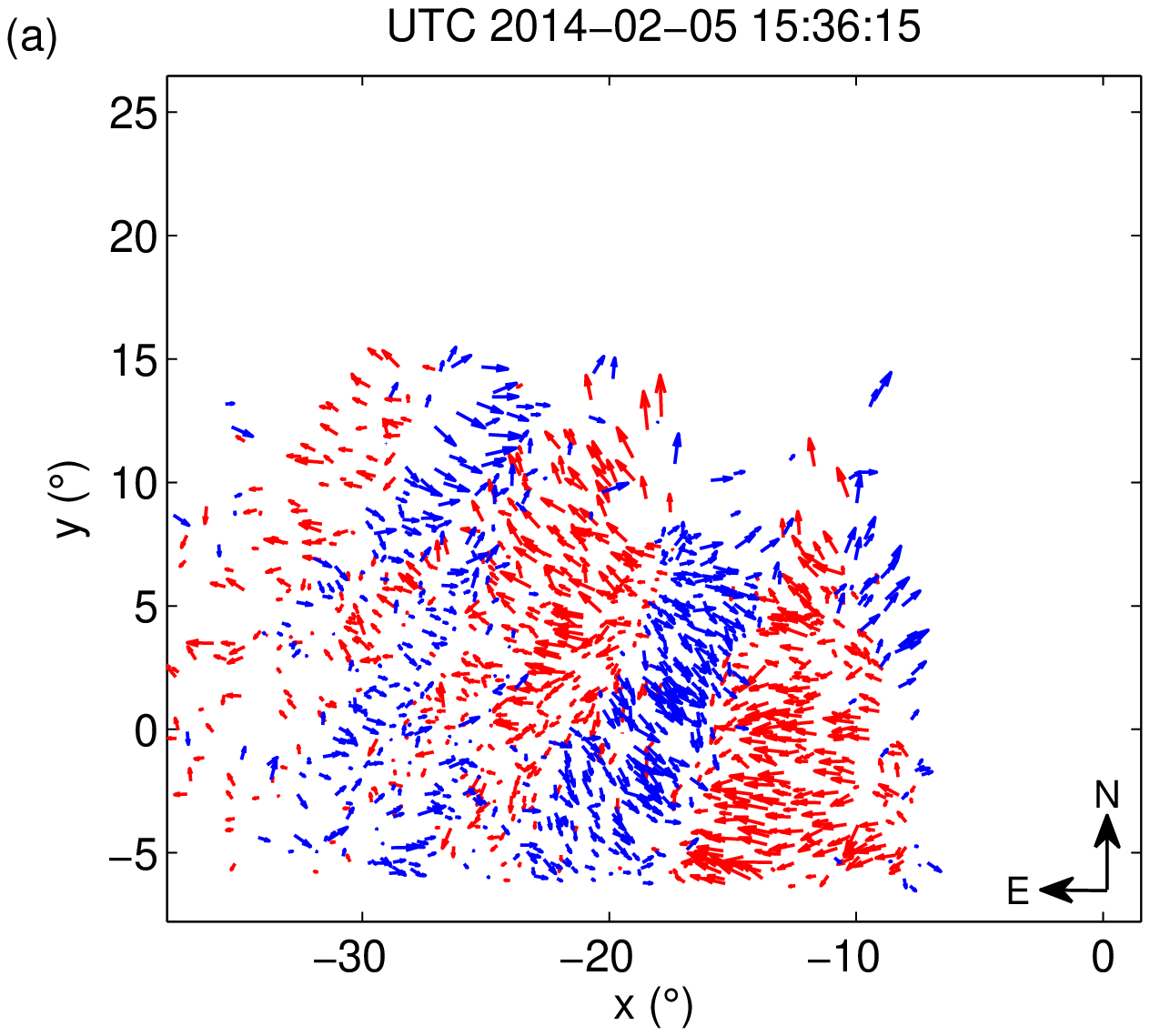}
  \includegraphics[width=0.49\textwidth]{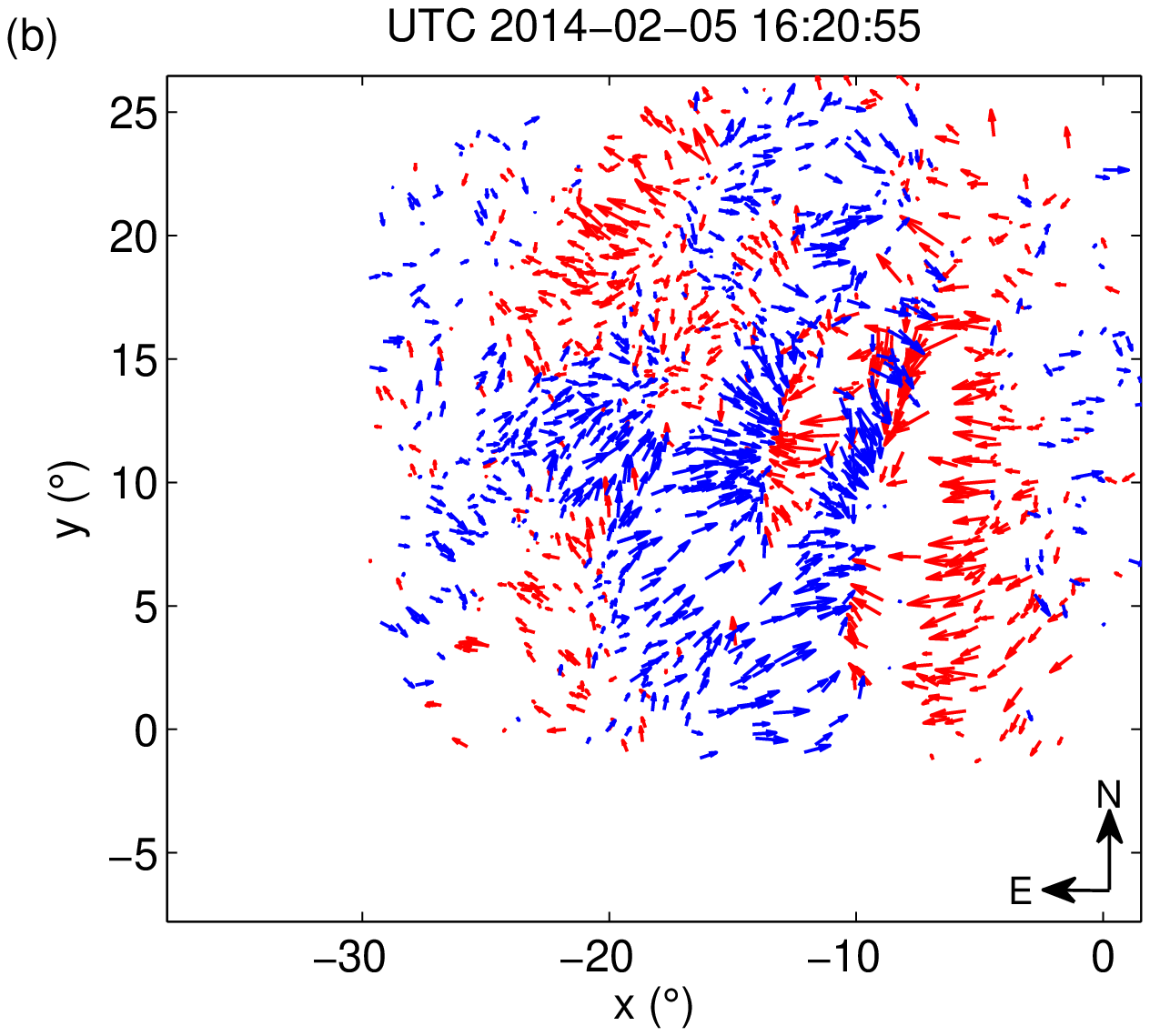}
  \caption{Spatial distribution of position offset vectors for two selected snapshots in dataset A, where the middle of each arrow marks the time-averaged location of a point source. The $x$-axis points west, the $y$-axis points north, and (0,0) marks the zenith. Arrows are colored red if they have a negative $x$-component (eastward-pointing), blue if their $x$-component is positive (westward-pointing). Arrow lengths are scaled to 100 times the actual displacement distance. A movie showing the time lapse of this vector field is included as supplementary material. The data are plotted using a Sanson-Flamsteed projection scheme (see \S\ref{sec:scalarfield} for details). See Movie S1 for an animation of the vector field for the remaining snapshots of this dataset.}
  \label{fig:arrowplots_A}
\end{figure}

Figure \ref{fig:arrowplots_A} shows the the position offset vector field for two selected snapshots from dataset A. There is a remarkable level of spatial organization into what appear to be patches of coordinated motion, where source displacements alternate between east and west on a spatial scale of $\sim$5$^\circ$ between reversals ($\sim$10$^\circ$ spatial periodicity). The patches are elongated roughly NW-SE (geographic). In the movie of the full dataset (see online supplementary material), these patches can be seen to drift slowly across the sky towards the NE.

In the case of dataset B, which was chosen for its relatively low level of ionospheric activity as inferred from the small magnitudes of position offsets compared to other MWA datasets, the position offset vector field displays isolated N-S elongated structures in the form of narrow vertical streaks (see Figure \ref{fig:arrowplots_B} for selected snapshots, and the corresponding movie in the supplementary material). These are much less pronounced than the bands in dataset A, and unlike in dataset A, there is no obvious periodicity in these structures. 

\begin{figure}[H]
  \centering
  \includegraphics[width=0.49\textwidth]{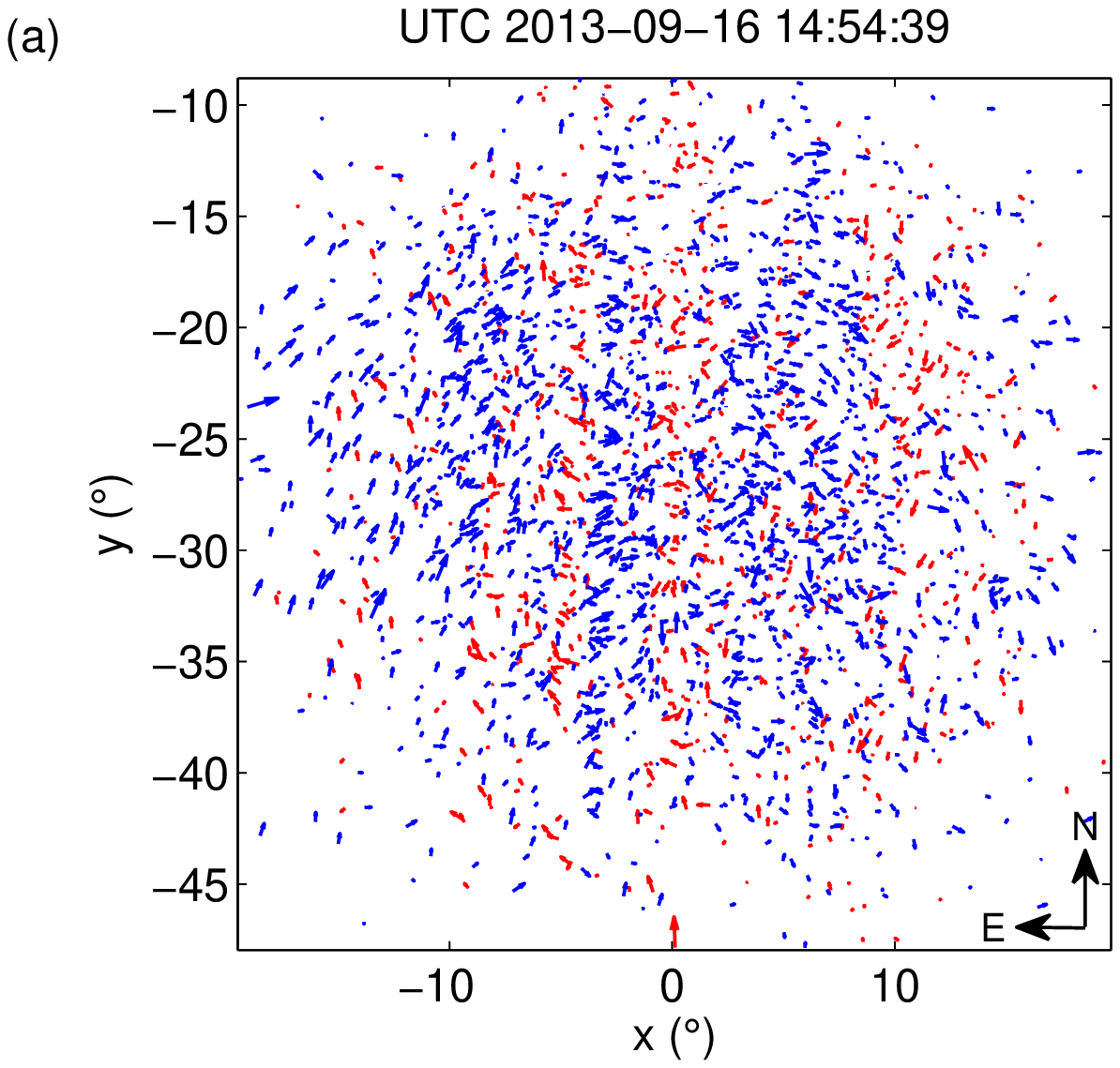}
  \includegraphics[width=0.49\textwidth]{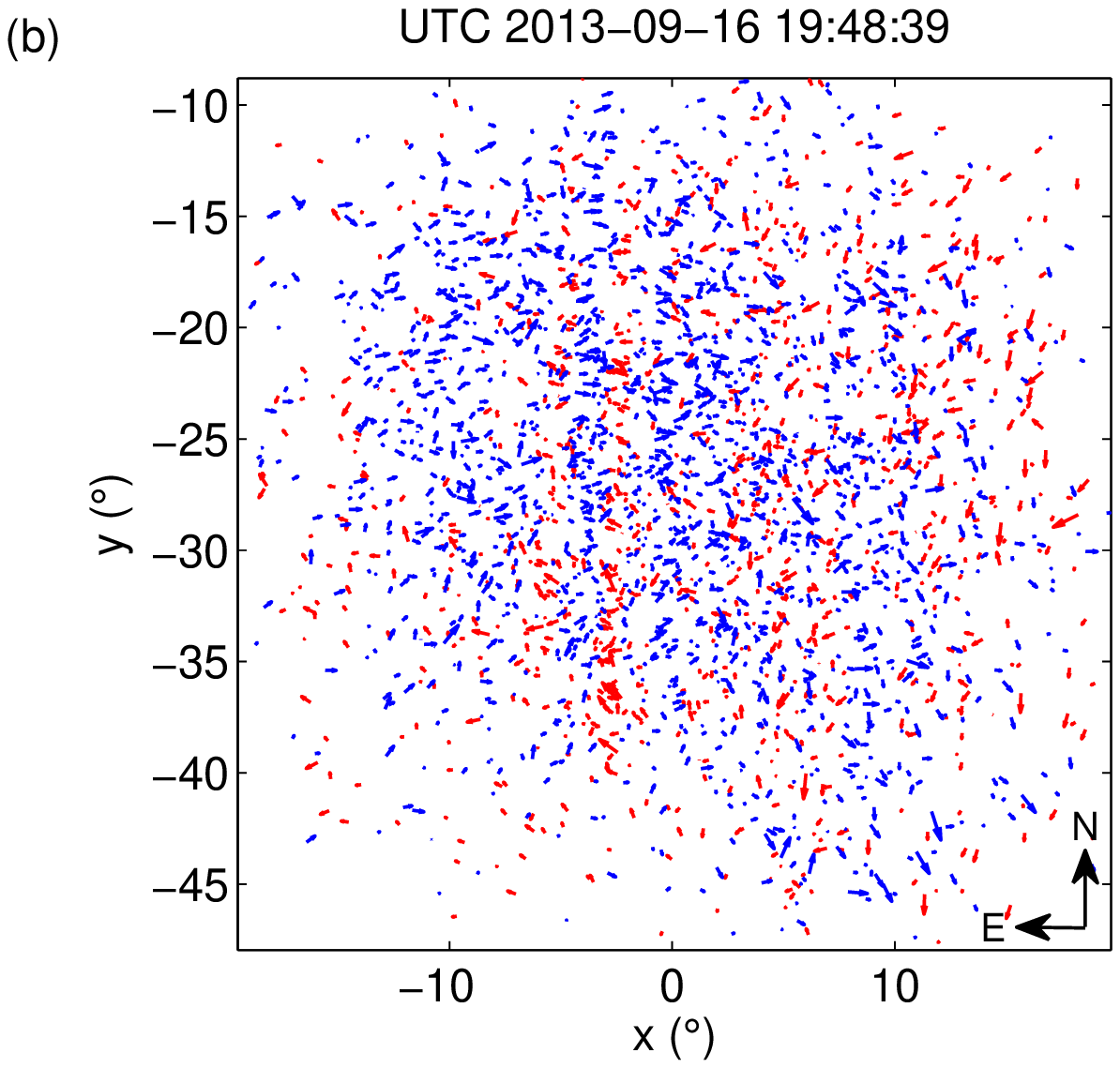}
  \caption{As for Figure \ref{fig:arrowplots_A}, but for dataset B and with arrow lengths scaled at 120 times the actual displacement distance. See Movie S2 for an animation of the vector field for the remaining snapshots of this dataset.}
  \label{fig:arrowplots_B}
\end{figure}

\begin{figure}[H]
  \centering
  \includegraphics[width=0.5\textwidth]{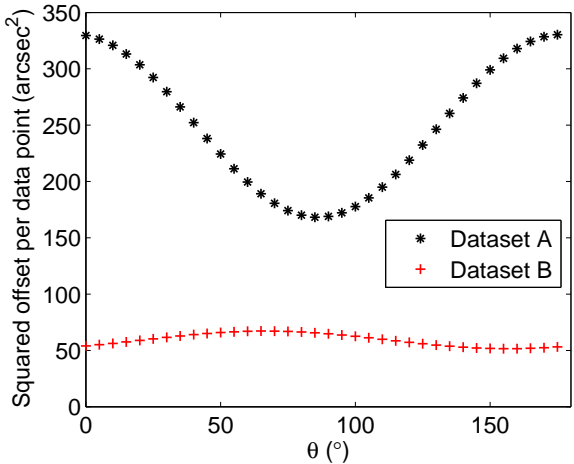}
  \caption{The total power $T(\theta)$ normalized to the number of measurement points in each dataset, as a function of the orientation of the axis of projection $\theta$, measured north of west.}
  \label{fig:totpower_vs_theta}
\end{figure}

The degree of anisotropy of a vector field can be quantified by examining how the average projected amplitude of the vectors varies as a function of $\theta$. The total power of fluctuations in each dataset is defined to be
\begin{equation}
  T(\theta) = \frac{V}{N_x N_y N_t} \sum_{\ell=1}^{N_x} \sum_{m=1}^{N_y} \sum_{n=1}^{N_t} \mathcal{P}_{\ell m n}(\theta) \label{eq:totpower}
\end{equation}
(integral of $\mathcal{P}(\theta)$ over all spatial and temporal frequencies, for a given projection angle $\theta$). The dependence of this on $\theta$ is plotted in Figure \ref{fig:totpower_vs_theta}. We see that the fluctuation power per pierce point is higher in dataset A than B, which is consistent with the visibly larger amplitude of motion in dataset A. 

Figure \ref{fig:totpower_vs_theta} shows that $T(\theta)$ maximizes along the E-W direction ($\theta \approx 0^\circ$) in dataset A, and along the NW-SE direction ($\theta \approx 65^\circ$) in dataset B. The act of projecting the vector field onto a given value of $\theta$ is analogous to passing light through a polarizing filter, and so we can define the degree of anisotropy (cf.~linear polarization) to be $(M-m)/(M+m)$, where $M$ and $m$ are the maximum and minimum values of $T(\theta)$, respectively. If the ionosphere is the main source of systematic anisotropies in the position offset vector field, then the degree of anisotropy is a nominal lower bound on the contribution of the ionosphere to the angular position offsets. The degree of anisotropy in dataset A is $\sim$30\%, and in dataset B it is $\sim$10\%. 

The power minima of Figure \ref{fig:totpower_vs_theta} correspond to characteristic residual position offsets of about 13\,arcsec for dataset A and 7\,arcsec for dataset B. These are upper bounds on the contribution of isotropic sources of error (e.g.~fitting errors, for roughly isotropic $u,v$-coverage) to the measured scatter in angular position. The mean value of the position fitting error $\Delta \psi$ (see Equation \ref{eq:poserror}) is around 3\,arcsec for both datasets, implying that fitting errors can only account for a small fraction of the position scatter. Assuming no other time-dependent effects, this indicates that a significant amount of the observed position shifts at 154 and 183\,MHz is due to the ionosphere.

\begin{figure}[H]
  \centering
  \includegraphics[width=0.5\textwidth]{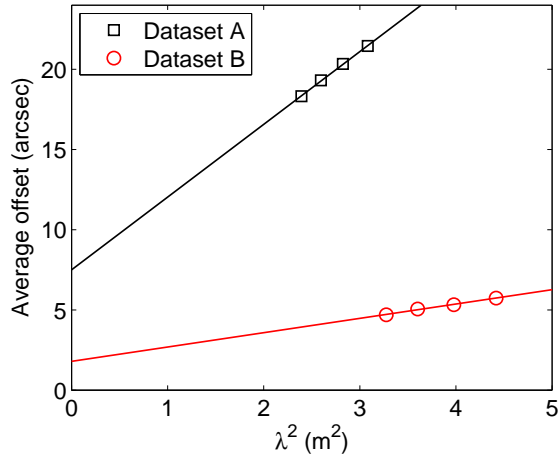}
  \caption{The magnitude of the angular position offset of sources in four frequency sub-bands, as a function of squared wavelength. Each data point was obtained by averaging the magnitude of offset over all sources appearing in that band, restricted to those brighter than 30$\sigma$. The lines are linear fits to the four data points. The error bars on each data point associated with random fitting errors are smaller than the marker size.}
  \label{fig:offset_vs_lambdasq}
\end{figure}

Figure \ref{fig:offset_vs_lambdasq} shows a linear dependence of position offsets on $\lambda^2$, where $\lambda$ is the observing wavelength, for bright ($\geq 30 \sigma$) sources. Such a dependence would be expected if the shifts were largely due to refractive effects (cf.~Equation (\ref{eq:offset})). However, the small fractional bandwidth of our observations prevents us from ruling out other functional dependencies. Although the dependence on $\lambda^2$ is well described by a straight line, the best-fit intercepts are non-zero. This suggests that multiple factors with different $\lambda$-dependences may be contributing to the position offsets, but a full investigation into this is beyond the scope of this paper.

In the following subsections we present the power spectra computed for the two datasets over a range of $\theta$ values. The power spectra for several representative values of $\theta$ are included in this paper. For other $\theta$ values, we refer the reader to supporting material comprising movies showing the evolution of the power spectra as the projection axis rotates through a full cycle. Since a unidirectional vector field vanishes when the axis of projection is orthogonal to it, by rotating the axis of projection we selectively filter out different wave modes and enhance others. In this manner, we infer the presence of multiple groups of waves propagating in a variety of directions and with different phase speeds. We verified with simulated data (see \S\ref{sec:pseudo1}) that our power spectra return results consistent with expectations.

\subsection{Response Function}\label{sec:response}
A rectilinear datacube of quasi-uniformly distributed points would be expected to produce sinc-like ringing in both space and time; however, the natural sampling pattern of the MWA is influenced by two additional factors. Firstly, the MWA primary beam response gradually decreases towards the edges of the FoV, leading to a smooth decline in the S/N ratio and thereby a decline in the density of sources. This tends to soften spatial ringing. Secondly, the diagonal tracks with respect to $(x,y,t)$ made by pierce points drifting across the sky are expected to produce diagonal ringing features with respect to $(k_x,k_y,\omega)$ in the response function, extending orthogonally to the drift direction in $(x,y,t)$. Since celestial sources drift westward (increasing $x$ with increasing $t$), we expect the ringing to be inclined towards negative $k_x$ with increasing $\omega$.

\begin{figure}[H]
  \centering
  \includegraphics[width=0.5\textwidth]{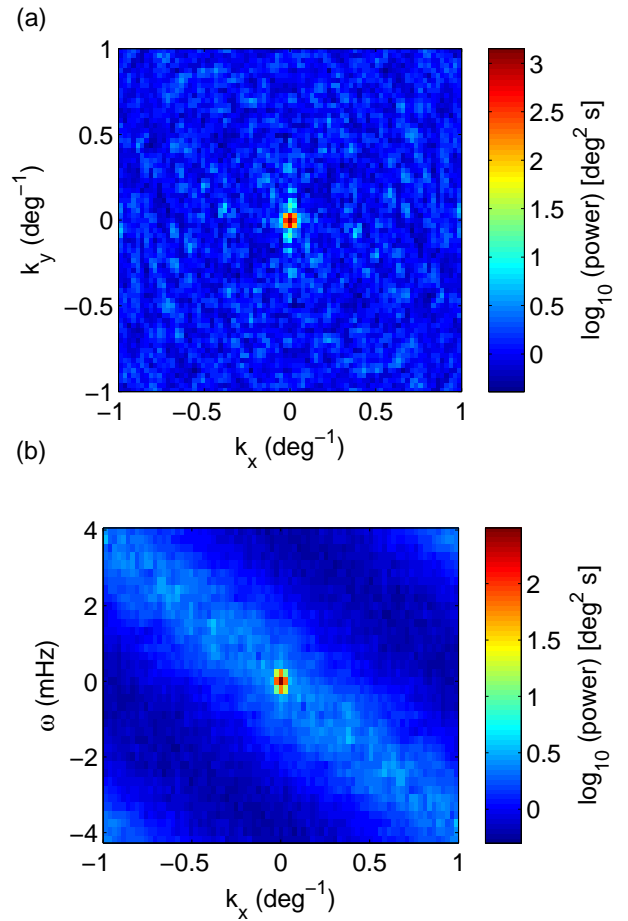}
  \caption{The response function of dataset A, where points have been gridded to a spatial resolution of 9 arcmin, averaged over (a) $\omega$ and (b) $k_y$.}
  \label{fig:EoR2_response}
\end{figure}

\begin{figure}[H]
  \centering
  \includegraphics[width=0.5\textwidth]{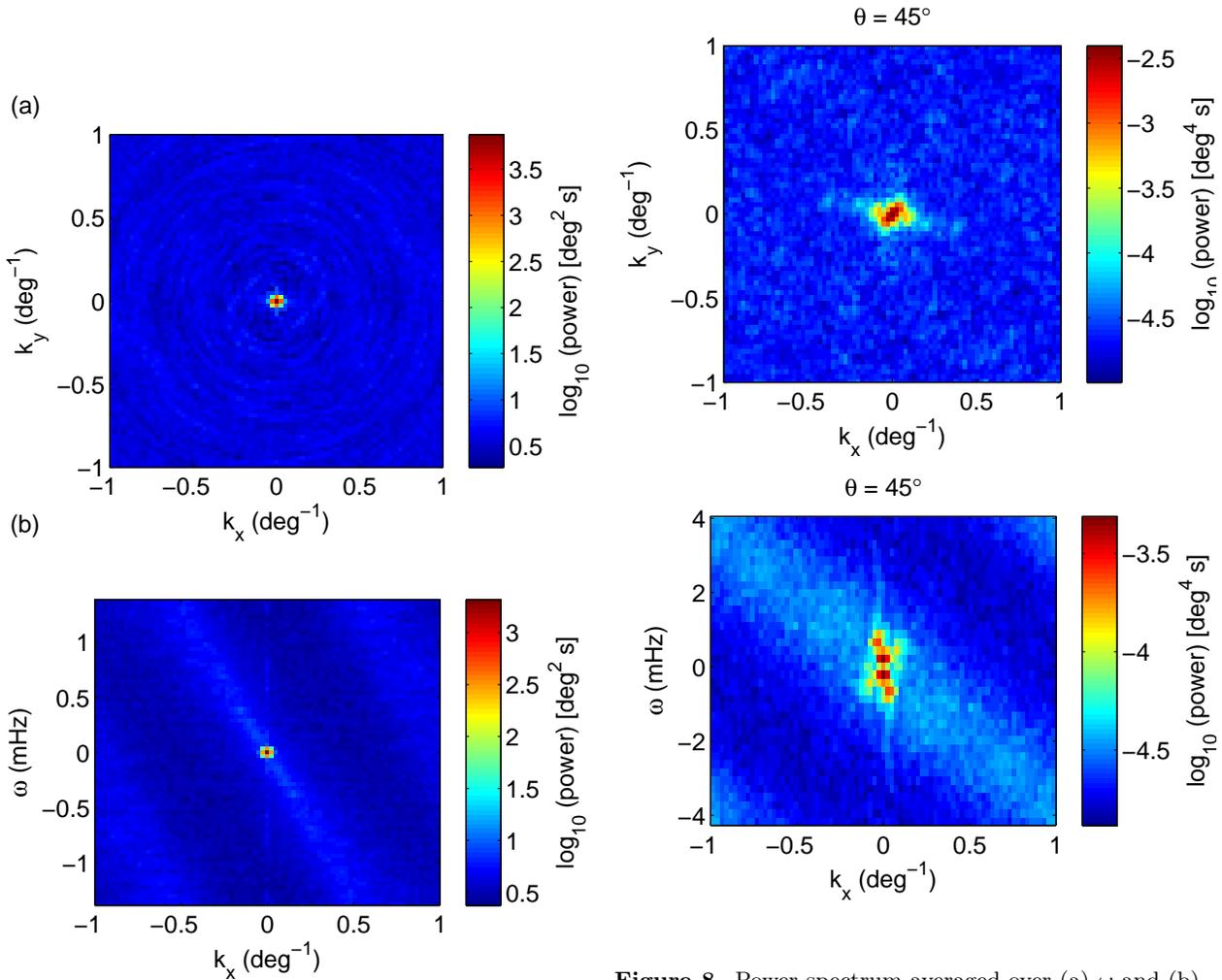}
  \caption{As for Figure \ref{fig:EoR2_response}, but for dataset B.}
  \label{fig:Sep-55_response}
\end{figure}

The response functions for datasets A and B after applying a Gaussian taper in the temporal dimension are shown in Figs \ref{fig:EoR2_response} and \ref{fig:Sep-55_response}. These each have a compact main lobe with minimal side lobes (note the logarithmic scale and $\delta$-function-like spike at the origin), which is desirable for spectral analysis, but there is evidence for the diagonal ringing patterns anticipated above. These features are more prominent in dataset A than B, and the main reason for this is the difference in survey mode between EoR and MWATS. Because the EoR survey tracks a fixed patch of celestial sky, the spacetime distribution of pierce points is concentrated about an axis that is skewed westward following the direction of source motion. Longer on-source times also imply that source tracks extend for greater distances on average than for MWATS. The drift speed of sources can in fact be calculated from Figure \ref{fig:EoR2_response}b as the gradient of the diagonal features: this is $\mathrm{d}\omega/\mathrm{d}k_x \approx 4 \times 10^{-3}$\,deg\,s$^{-1}$, which agrees with the angular rotation speed of the Earth at the latitude of the MWA (360$^\circ \cos 26\fdg7$\,day$^{-1} = 3.7 \times 10^{-3}$\,deg\,s$^{-1}$).

\subsection{Power Spectra for Dataset A} \label{sec:spectraA}
\begin{figure}[H]
  \centering
  \includegraphics[width=0.5\textwidth]{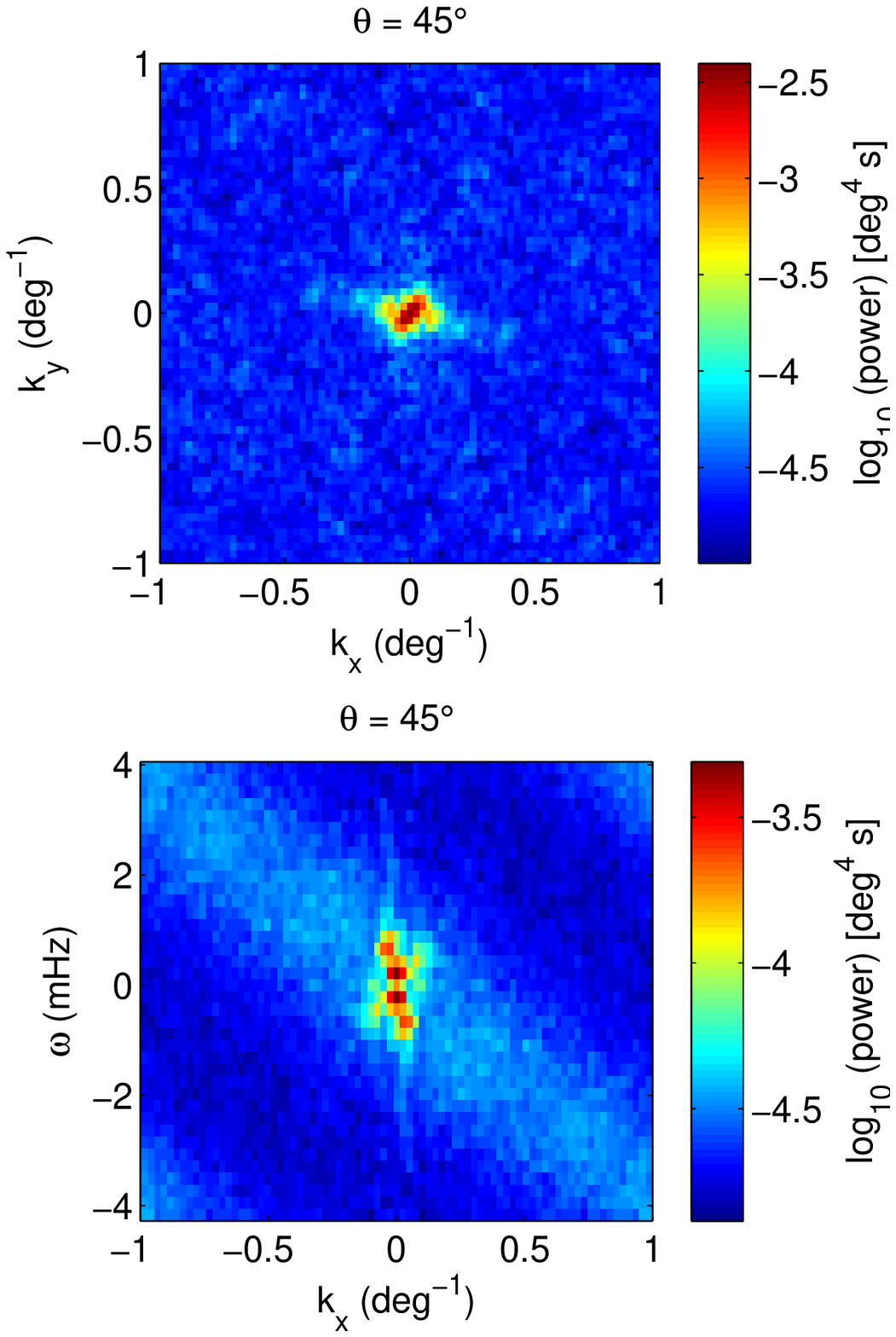}
  \caption{Power spectrum averaged over (a) $\omega$ and (b) $k_y$ for dataset A, for $\theta = 45^\circ$. See Movie S3 for an animation of the power spectrum for this dataset as the projection axis rotates.}
  \label{fig:EoR2_theta45}
\end{figure}

\begin{figure}[H]
  \centering
  \includegraphics[width=0.5\textwidth]{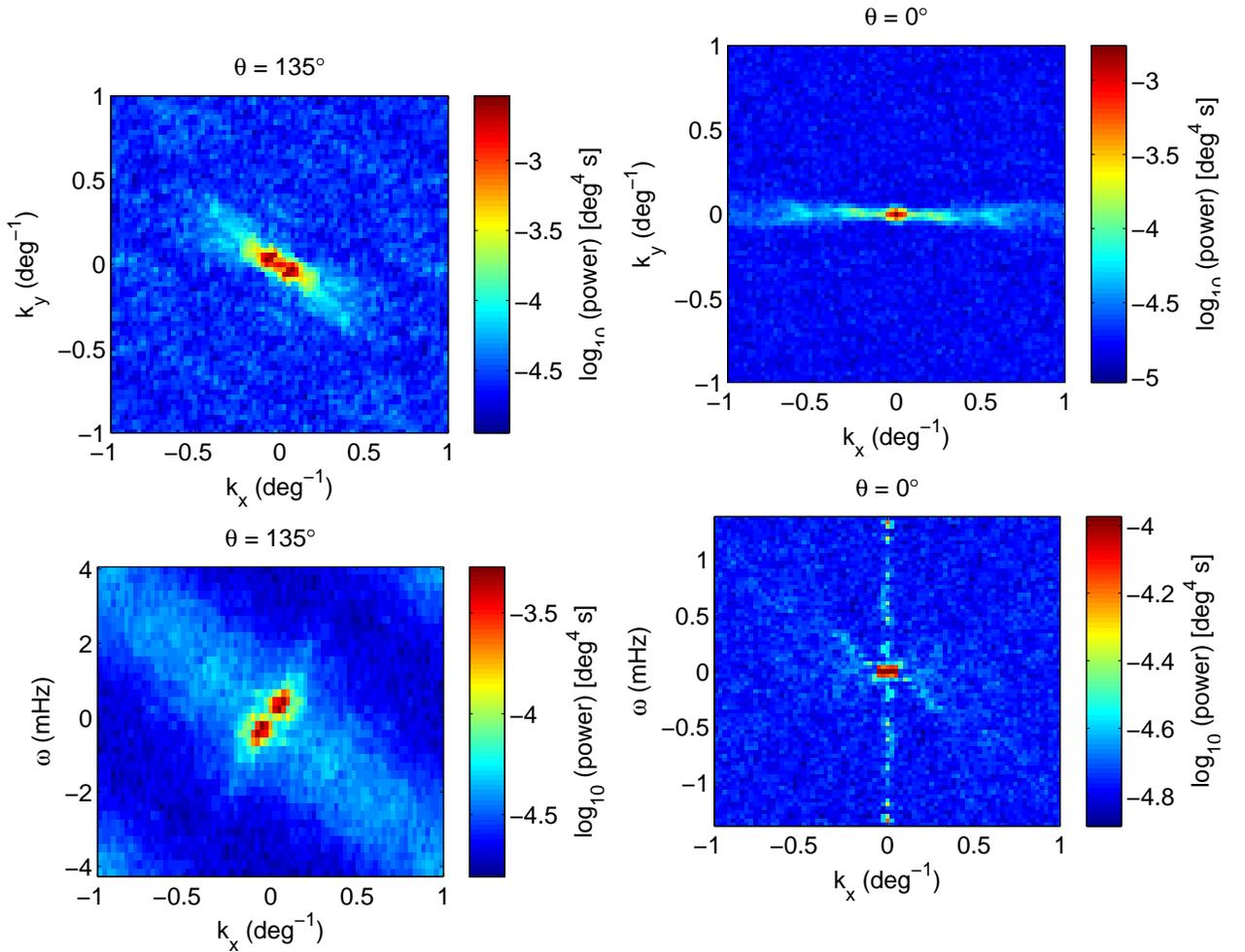}
  \caption{As for Figure \ref{fig:EoR2_theta45}, but with $\theta = 135^\circ$.}
  \label{fig:EoR2_theta135}
\end{figure}

The power spectra for $\theta = 45^\circ$ and 135$^\circ$ are shown in Figs \ref{fig:EoR2_theta45} and \ref{fig:EoR2_theta135}, respectively. The dominant mode for $\theta = 45^\circ$ is a low spatial-frequency mode with a period of $\sim$1\,hr, where the power spectral density distribution is elongated along the direction $50^\circ$ anti-clockwise from the positive $k_x$ axis. This corresponds to large-scale structure (angular scale of order the FoV) with phase fronts aligned roughly NE-SW travelling towards the NW. We discuss this in \S\ref{sec:fastlarge}.

For $\theta = 135^\circ$ there are strong peaks at $|k| \approx 0.1$\,deg$^{-1}$, corresponding to phase fronts spaced regularly by $\sim 10^\circ$ and stretching along $\theta \approx 50^\circ$ (i.e.~NW-SE, geographic). This matches the direction of the elongation seen for the bands in Figure \ref{fig:arrowplots_A} and their spacing as estimated by eye. The lower panel in Figure \ref{fig:EoR2_theta135} reveals the temporal frequency to be about 0.4\,mHz (period of 40\,min). The corresponding phase velocity is about $4 \times 10^{-3}$\,deg\,s$^{-1}$ towards the NE. There is additional power spreading out to higher spatial frequencies, indicating the presence of smaller-scale structure up to $|k| \approx 0.5$\,deg$^{-1}$ (angular scales of $\sim 2^\circ$). We discuss these observations further in \S\ref{sec:slowsmall}.

\subsection{Power Spectra for Dataset B} \label{sec:spectraB}
\begin{figure}[H]
  \centering
  \includegraphics[width=0.5\textwidth]{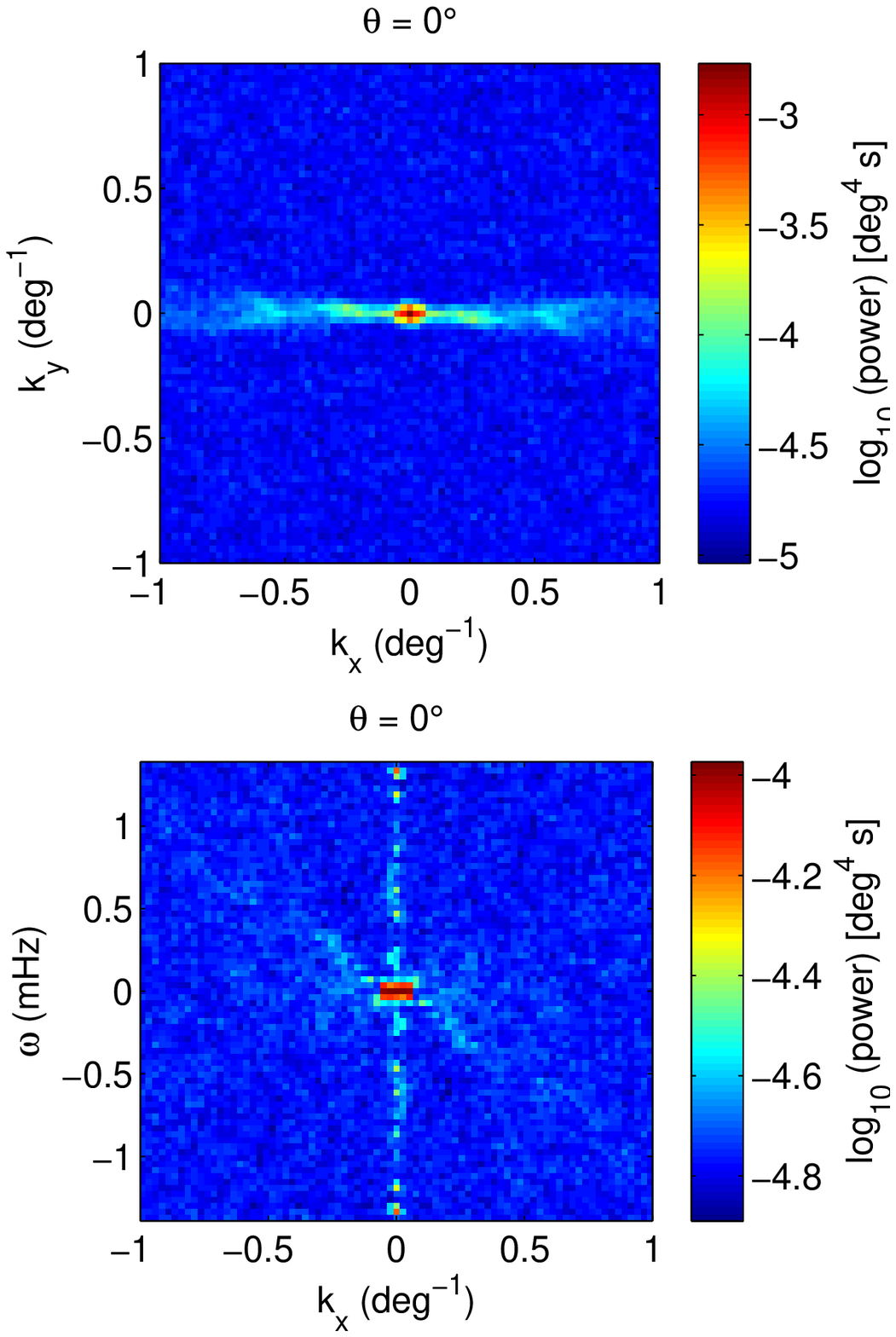}
  \caption{As for Figure \ref{fig:EoR2_theta45}, but for dataset B with $\theta = 0^\circ$. See Movie S4 for an animation of the power spectrum for this dataset as the projection axis rotates.}
  \label{fig:Sep-55_theta0}
\end{figure}

\begin{figure}[H]
  \centering
  \includegraphics[width=0.5\textwidth]{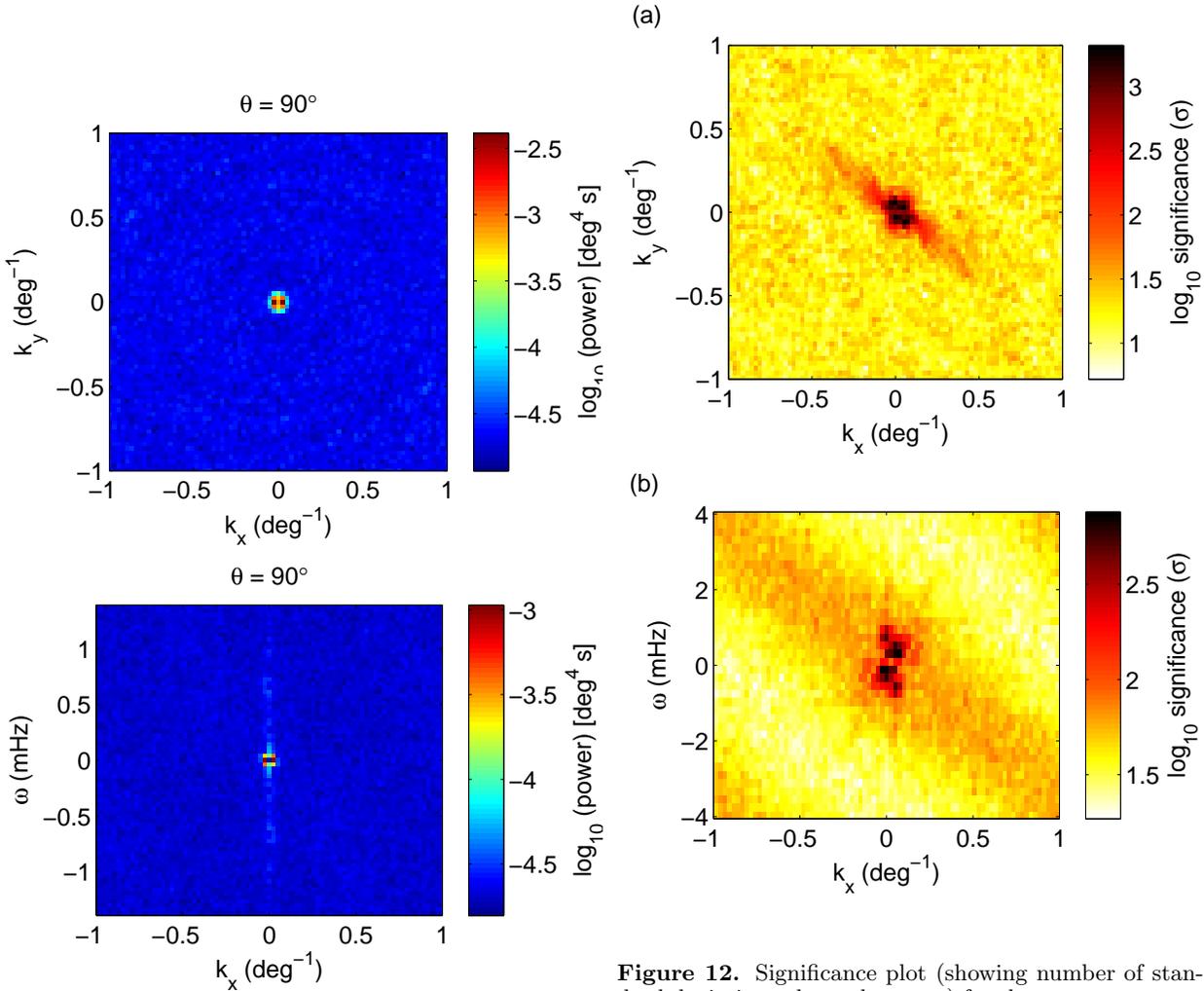}
  \caption{As for Figure \ref{fig:Sep-55_theta0}, but with $\theta = 90^\circ$.}
  \label{fig:Sep-55_theta90}
\end{figure}

The power spectra for $\theta = 0^\circ$ and 90$^\circ$ are shown in Figs \ref{fig:Sep-55_theta0} and \ref{fig:Sep-55_theta90}, respectively. In Figure \ref{fig:Sep-55_theta0}, the power spectral density distribution is concentrated along $k_y = 0$ over a substantial range of $k_x$, out to $|k_x| = 1$\,deg$^{-1}$. This indicates the presence of small-scale structure with angular scales of $\sim 1^\circ$ highly elongated along the N-S direction, but with no strong periodicity. The lower panel ($\omega,k_x$-plane) shows no clear concentration of power associated with higher $k_x$ modes, meaning that these small-scale structures do not drift with a well-defined phase velocity. We discuss these in \S\ref{sec:slowsmall}.

There is a peak at $\omega = 0$ and the lowest non-zero $k_x$ value probed by the MWA that can be seen in Figure \ref{fig:Sep-55_theta90}, indicating the presence of a stationary mode with a wavelength of order the size of the FoV. We discuss possible explanations for this feature in \S\ref{sec:stationary}.

For $\theta = 135^\circ$ there appears to be a peak at around $\omega = 0.6$\,mHz near $|k| = 0$, corresponding to a long-wavelength fluctuation of angular width of order the FoV with a period of about 30\,min. However, we are unable to determine the propagation direction because this is not resolved in the $k_x,k_y$-plane. We discuss this in \S\ref{sec:fastlarge}.

\subsection{Error analysis}
\begin{figure}[H]
  \centering
  \includegraphics[width=0.5\textwidth]{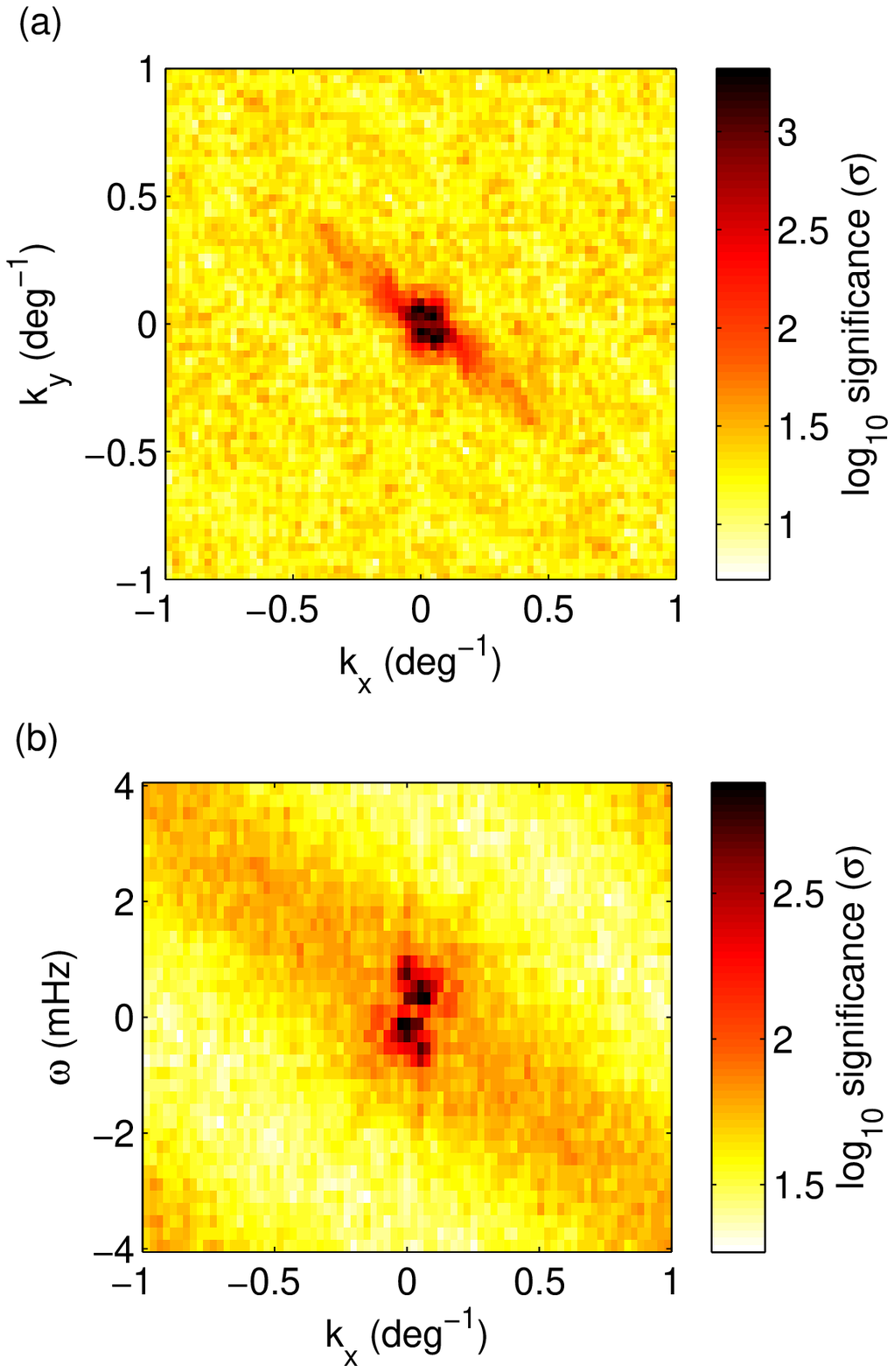}
  \caption{Significance plot (showing number of standard deviations above the mean) for the power spectrum of dataset A, averaged over (a) $\omega$ and (b) $k_y$, for $\theta = 90^\circ$, on a logarithmic scale. This was generated using 200 Monte Carlo iterations.}
  \label{fig:EoR2_significance}
\end{figure}

We performed an error analysis by comparing the amplitudes of the features in the power spectra with the noise level expected from position fitting errors. We used a Monte Carlo approach where we sampled position offsets from Gaussian distributions with characteristic widths given by Equation (\ref{eq:poserror}), and then computed the power spectra. We took the significance of the data to be the number of standard deviations above the mean, where the standard deviations and means were computed for each spatiotemporal frequency over 200 Monte Carlo iterations. All features in the power spectra are many standard deviations above the mean, indicating that the measured signals are significantly larger than expected of random fluctuations. Figure \ref{fig:EoR2_significance} shows an example of a significance plot for dataset A, where the quantity plotted is the number of standard deviations above the mean.

\begin{figure}[H]
  \centering
  \includegraphics[width=0.5\textwidth]{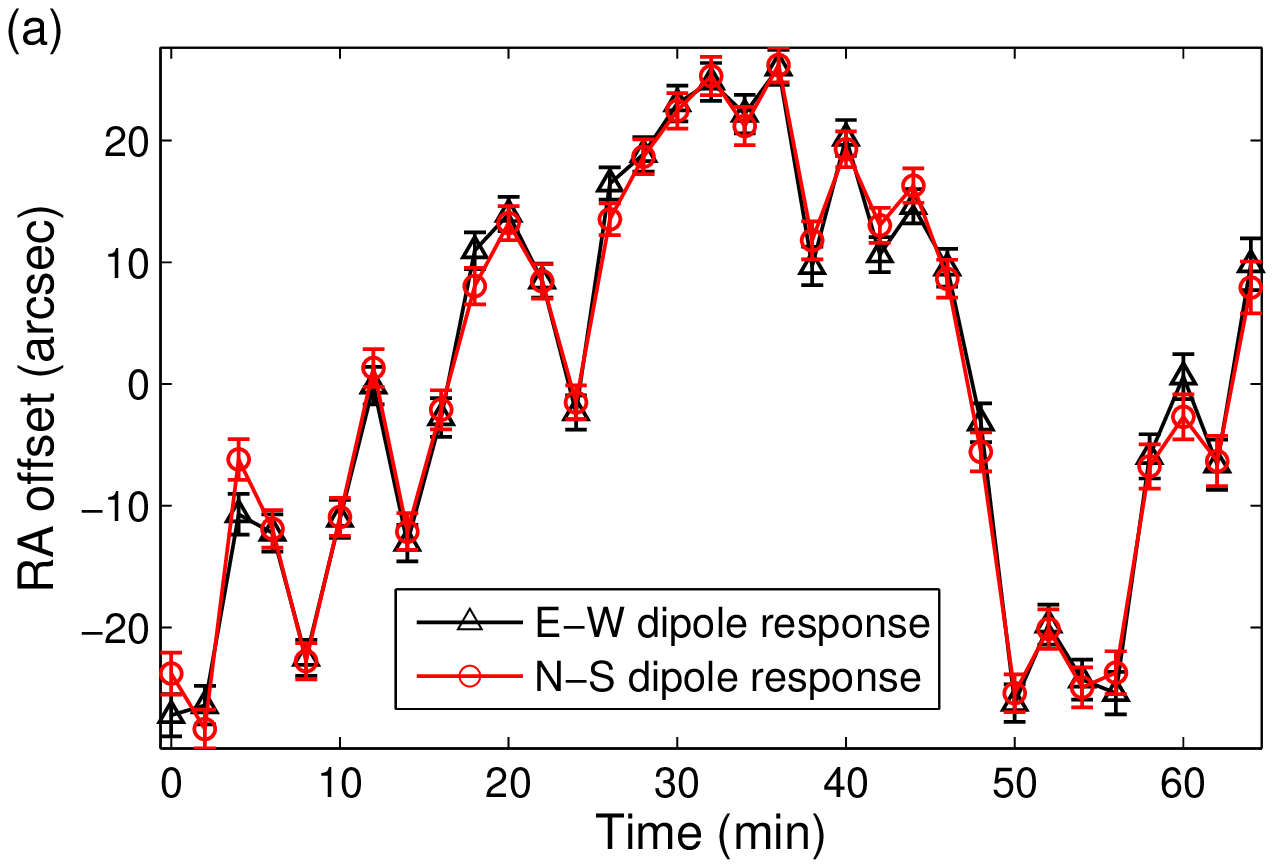}
  \includegraphics[width=0.5\textwidth]{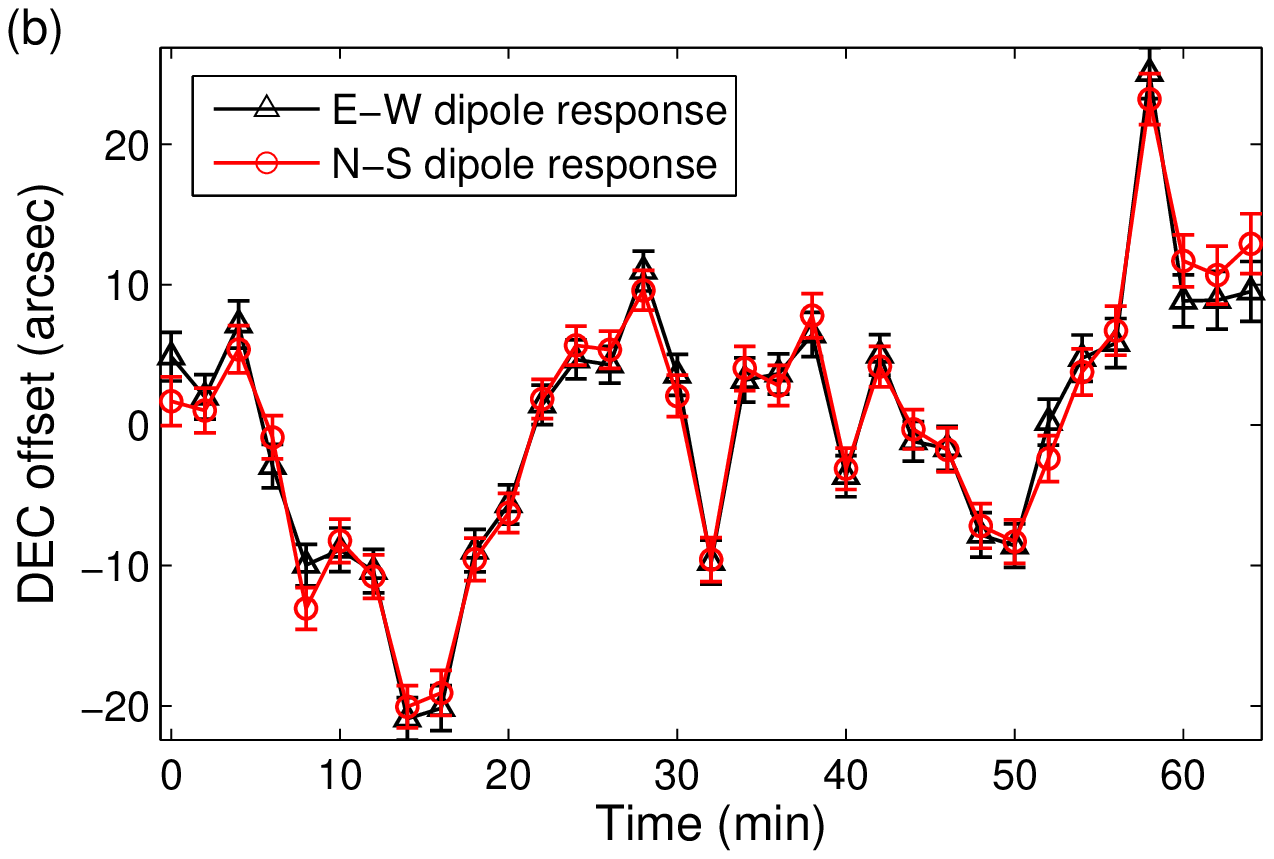}
  \caption{Comparison of the angular offsets of a bright (35-$\sigma$) source between images recorded in the two independent instrumental polarizations, along the (a) RA and (b) Dec directions (these are orthogonal). Error bars indicate the precision associated with Gaussian fitting. The two independent measurements are consistent with one another to within the measurement precision, which is significantly smaller than the amplitude of the position fluctuations themselves.}
  \label{fig:XXvsYY_offset}
\end{figure}

An alternative means of estimating the error is to compare the measured positions between images generated from the two instrumental polarizations (the N-S and E-W dipole responses), since these are effectively two independent, simultaneous measurements of the same signal. The ionosphere should have identical effects on these images, and so any discrepancies represent the contribution of random noise. We find that the two corresponding sets of position measurements agree to within the position fitting errors predicted by Equation (\ref{eq:poserror}). An example comparison between the two polarization responses for a bright radio source is shown in Figure \ref{fig:XXvsYY_offset}, where the position offsets for the two independent measurements are seen to be consistent within error bars.

Although it is reassuring that the signals are strong compared to the amplitude of noise expected from fitting errors, the most problematic sources of error are likely to be those that are coherent rather than incoherent. The power spectral density of incoherent noise distributes itself fairly evenly over the various frequency modes, and so should not interfere with the identification of coherent signals, which are localized in Fourier space. However, spurious structure or the distortion of features in the power spectra can arise from a variety of causes, including sidelobes of the response functions (see Figs \ref{fig:EoR2_response} and \ref{fig:Sep-55_response}), spectral leakage, and imaging artefacts. 

\section{Discussion}\label{sec:discussion}
Similar power spectrum analyses of point-source position offsets have been performed previously by \citet{Helmboldt2012b} and \citet{Helmboldt2012c} using the VLA. However, the spatial sampling of our data is 1--2 orders of magnitude more complete and our FoV several times larger, which has not only yielded a better-behaved response function for power spectrum analysis, but has also allowed us to reconstruct ionospheric waveforms directly from the images. The MWA is the first radio telescope with the ability to do this; in previous studies making use of radio interferometers \citep{Jacobson1993, Jacobson1996, Hoogeveen1997a, Hoogeveen1997, Helmboldt2014a}, the spatial sampling of the ionosphere was too sparse to visualize the density structures in any detail.

We have identified several types of fluctuations in the MWA data occupying different regions of Fourier space. These include the spatially-resolved, slowly-drifting fluctuations visible in Figs \ref{fig:arrowplots_A} and \ref{fig:arrowplots_B}, and large-scale, unresolved fluctuations near $|k| = 0$. Of the large-scale fluctuations, we have identified those with well-defined periods of 30\,min and 1\,hr, and also those that appear to be stationary with respect to the FoV ($\omega = 0$).

\subsection{Slowly-Drifting, Small-Scale Fluctuations}\label{sec:slowsmall}
As described in \S\ref{sec:results}, spatially-resolved fluctuations were detected in both datasets A and B, where the structures appear to be elongated greatly along a certain direction. Comparison with the Earth's magnetic field parameters at the location of the MWA (magnetic inclination of $-60^\circ$, magnetic declination of $0^\circ$) reveals that these are consistent with being irregularities that are extended along the geomagnetic field lines. Their drifts where measured ($4 \times 10^{-3}$\,deg\,s$^{-1}$ in dataset A) are several times slower than would be expected of TID propagation speeds at thermospheric heights ($\gtrsim$100\,m\,s$^{-1}$, or $\gtrsim$0.02\,deg\,s$^{-1}$ at 300\,km altitude). This can be explained if they are in-situ stationary structures that are drifting along with the background plasma. Drift speeds of several tens of meters per second are expected for neutral wind motions at thermospheric heights \citep{Amayenc1972}.

At altitudes above about 150\,km, where the electrical conductivity along the Earth's magnetic field lines is substantially higher than transverse to them, the field lines can be treated as equipotentials \citep{Park1973}. At these heights, diffusion along the field lines is much faster than across them, and density enhancements that may initially form along a localized portion of the field lines readily diffuse along them to form large-scale, duct-like structures. These may then persist for significant durations, of order the timescales governing cross-field diffusion, which can be hours to days \citep{Thomson1978}.

Although we have not estimated the altitudes of the structures seen here, previous analyses of field-aligned density ducts in MWA data by \citet{Loi2015} have obtained values of $\sim$400--1000\,km. If the structures lie at around 600\,km altitude, their physical widths would be between 10--100\,km, similar to values measured by satellites for field-aligned ducts \citep{Angerami1970} and the magnetic eastward-directed plasmaspheric waves described by \citet{Helmboldt2012b}. The physical drift speeds in dataset A would be about 40\,m\,s$^{-1}$ at this altitude. If the local fractional density variations are comparable between the two datasets, the greater prominence of the structures in dataset A than dataset B can be explained if they are at a lower altitude in the former, since background densities are larger at lower altitudes. The suggestion that the structures in dataset A are nearer the ground, if the physical widths are intrinsically similar between the two datasets, is consistent with the larger angular widths observed in dataset A.

Solar activity was significantly higher during the interval of dataset A (10.7-cm radio flux index F10.7 = 188 sfu, where 1 sfu = $10^{-22}$\,W\,m$^{-2}$\,Hz$^{-1}$) compared to dataset B (F10.7 = 93 sfu). However, similarly extensive structuring is also seen in MWA data during periods of low solar activity, and so this appears not to be a decisive influence on the observed behaviour. There is no clear link to geomagnetic activity that may explain the differences between the two datasets: both were obtained under quiet conditions where the $K_p$ global storm index (quantifying fluctuations of the Earth's magnetic field) was around 1, the 24-hr maximum $K_p$ index not exceeding 2 on either day. A broad study performed by \citet{Hoogeveen1997a} using the Los Alamos radio interferometer found that field-aligned ducts of similar description appear most prominently on nights just after geomagnetic storms. Although we have yet to examine MWA data taken under significantly disturbed geomagnetic conditions, we have noted that field-aligned ducts appear routinely under quiet to mild conditions. Global GPS TEC maps \citep[e.g.][]{Rideout2006} have insufficient TEC precision and spatial resolution to reveal structures on these scales, but detections may still be possible with single-receiver relative TEC measurements. A climatological study of ionospheric behaviour above the observatory making use of combined MWA and on-site GPS TEC measurements is the subject of ongoing work.

The steeply inclined nature inferred for the density structures (60$^\circ$ to the ground, as given by the magnetic inclination at the MRO) is at odds with the constant-altitude, thin-screen approximations commonly used to model the ionosphere and employed in some calibration strategies \citep{Rino1982, Wernik2007, Intema2009}. So far, no one attempting screen-based calibration has accounted for the possibility that density irregularities that affect radio observations may be localized to \textit{magnetic shells} rather than constant-altitude layers. The Earth's magnetic field is well described by a dipole, and so the field lines are generally inclined with respect to the ground: they are horizontal at the magnetic equator, vertical at the magnetic poles, and at intermediate inclinations at other latitudes.

Although the bulk of the TEC is indeed concentrated in a relatively thin layer at constant altitude, the constant offset component of the TEC is invisible to a radio interferometer, since this adds a constant phase to all antennas which then cancels out. Interferometers differ fundamentally from conventional techniques for probing the ionosphere, the latter of which are mostly sensitive to absolute density and therefore tend to be dominated by the signal from the layer of peak electron density (300--400\,km altitude). Because an interferometer is sensitive to $\nabla_\perp$ TEC rather than absolute TEC, it experiences no such bias: irregularities at a large range of altitudes can easily dominate the signal, including those well above the peak density layer.

The physically compact size of the MWA simplifies the problem of ionospheric calibration by minimizing the differences between the phase screens seen by different antennas \citep[cf.][]{Lonsdale2005}, thus reducing the dimensionality of the calibration problem to simply a function of angular position (besides time). However, screen calibration models that may eventually be implemented on larger arrays (e.g.~LOFAR) may need to allow for inclined screen components, to capture this tendency for magnetic shell localization. We find that duct-like, field-aligned (and therefore steeply inclined) structures are very common in MWA data, appearing in half or more of all nights examined. This implies that a large population of irregularities causing measurable distortions in MWA data are ones that cannot be captured by a constant-altitude screen model.

\subsection{Large-Scale, Short-Period Fluctuations}\label{sec:fastlarge}
The long-wavelength ($|k| \approx 0$), 30\,min-period fluctuations detected in dataset B are likely to be associated with a medium-scale TID (MSTID). MSTIDs have wavelengths of several hundred kilometers (of order the FoV or larger, at 300\,km altitude) and periods of several tens of minutes \citep{Hunsucker1982}, so these give rise to fluctuations occupying low-spatial frequency, high-temporal frequency regions of Fourier space. These are the most common type of ionospheric fluctuation believed to occur at mid-latitudes. We detect their spatiotemporal signatures in many other MWA datasets.

The large-scale, time-resolved fluctuation in dataset A with a measured period of about 1\,hr could be a large-scale TID (LSTID). LSTIDs have wavelengths of $\sim$1000\,km and periods of an hour or more \citep{Georges1968}. However, noting that its inferred propagation direction is parallel to the field lines, this may not in fact be a fluctuation confined to the thermosphere (as is the case for TIDs) but instead represent the wave-like transport of material up or down inclined magnetic flux tubes. While LSTIDs are expected to produce periodic, quasi-sinusoidal oscillations, the relatively short duration of our observation for dataset A does not allow us to distinguish between this and a one-off wave of deposition. Flows between the plasmasphere and ionosphere are expected to occur at night, when decay of the ionosphere via recombination is combatted by downflows from the plasmaspheric reservoir \citep{Park1970, Helmboldt2012}.

There is a possibility that oscillating disturbances such as TIDs may be aliased at MWA survey cadences. The shortest TID wave period measurable at the cadence of the two MWA surveys used here is 4\,min for dataset A and 12\,min for dataset B. Gravity waves, which comprise MSTIDs and LSTIDs, can only propagate at frequencies below the local Brunt-V\"{a}is\"{a}l\"{a} (B-V) frequency, also known as the buoyancy frequency \citep{Yeh1974}. The B-V frequency at ionospheric heights corresponds to a period of 10--12\,min \citep{Hines1960}, and so in the absence of background winds these cadences should be adequate for characterizing MSTIDs and LSTIDs. However, background winds can Doppler-shift the waves to higher frequencies exceeding the Nyquist limit. The presence of winds with speeds as high as 100\,m\,s$^{-1}$ has been the proposed explanation for observations of gravity waves with periods as short as 5\,min \citep{Davies1973}. Small-scale TIDs (SSTIDs), which are acoustic waves rather than gravity waves \citep{Hunsucker1982}, oscillate with frequencies above the B-V frequency and may be aliased at these cadences. However, raw interferometer visibilities are often stored at 0.5\,s resolution, meaning that the data can be re-imaged under a much higher time resolution if necessary.

\subsection{Large-Scale, Stationary Fluctuations}\label{sec:stationary}
The peak at $\omega = 0$ and the lowest non-zero spatial frequency in dataset B implies the existence of a fluctuation that is stationary over the FoV, with a length scale of order the FoV. This can be explained by way of either ionospheric or instrumental effects. Both are likely to contribute at some level, but we have not investigated into which may be the more dominant cause. 

A completely still, uniform ionosphere at a constant altitude can produce a systematic distortion in the angular position shifts of background sources because of the $\text{sec} \zeta$ dependence of the TEC on the zenith angle $\zeta$. The MWA FoV is about 30$^\circ$ wide, which implies that for a zenith pointing, the TEC along lines of sight near the edge of the FoV is about 4\% larger than at zenith. For a background TEC of order 10\,TECU (where 1\,TECU = $10^{16}$\,electrons\,m$^{-2}$) and an ionospheric altitude of 300\,km, the associated radial gradient is about $4 \times 10^{10}$\,m$^{-3}$, which would produce position shifts of order 10\,arcsec at 154\,MHz. This is larger than typical position fitting errors and should be detectable via power spectrum analysis, since this signal would be coherent and persistent. Sources refract towards regions of lower TEC, and so the offset vectors associated with this effect would be radially inward.

Although this ionospheric distortion should be circularly symmetric about the zenith, the signal appearing in the power spectra for dataset B is concentrated along E-W. This can be explained by the fact that the time-averaged position of each source is taken as the reference position. This causes stationary features with respect to the sources to be subtracted out (recall the discussion in \S\ref{sec:scalarfield}). Since celestial sources drift E-W, N-S components of the distortion pattern would be removed, leaving just an E-W signal.

Instrumental effects that may give rise to large-scale distortions of source angular positions include tile position errors, and errors in correcting for the $w$-terms during imaging \citep[these describe the deviation of the array from coplanarity;][]{Perley1999}. Unlike the $\text{sec}\zeta$ distortion produced by the ionosphere, $w$-correction errors can give rise to distortion patterns that are radially inward, outward, and/or with azimuthal components. 

Applying a low-pass spatial filter to dataset B reveals that the distortion pattern associated with the $\omega = 0$ peak is radially inward, potentially consistent with either source of error. However, in other MWATS datasets where $\omega = 0$ peaks appear, similar checks sometimes reveal distortions that are radially outward. This opposes the direction expected for ionospheric distortions, indicating that systematic angular distortions may exist in MWA data that are not necessarily ionospheric in origin. 

\subsection{Turbulence}\label{sec:turbulence}
Kolmogorov turbulence \citep{Kolmogorov1941} is expected to produce a power spectrum that falls off as $k^{-5/3}$. Note that although the 3D power spectrum of TEC fluctuations goes as $k^{-11/3}$, the quantity measured here is the TEC gradient. Taking a derivative multiplies the fluctuation spectrum by a factor of $i|k|$ and the power spectrum by a factor of $k^2$, leading to an overall $k^{-5/3}$ dependence \citep{Helmboldt2012}. The broad distribution of power spectral density along $k_y = 0$ in the top panel of Figure \ref{fig:Sep-55_theta0} may be evidence for a turbulent cascade, but this is not the only possibility. Narrow N-S density structures can also produce a power spectral density distribution that is broadly extended along $k_y = 0$ (cf.~Fourier transform of a $\delta$-function). That narrow structures indeed appear in dataset B (Figure \ref{fig:arrowplots_B}) suggests that these are the more likely explanation for the observed power along $k_y = 0$, rather than an anisotropic turbulent spectrum, although we cannot rule out the existence of the latter.

A further complication arises from the effects of spectral leakage and noise from incoherent measurement errors, both of which decrease the S/N ratio in the power spectra. These would flatten the power spectral density and preclude reliable measurement of the power-law index of a turbulent component, should there be one present (see \S\ref{sec:pseudo2}). A proper quantification of all major sources of error and deconvolution to remove sampling artefacts will be required to measure the spatiotemporal properties of TEC fluctuations with greater accuracy. 

\section{Conclusions \& Outlook}\label{sec:conclusions}
We have performed the first power spectrum analysis of ionospheric TEC variations using the MWA. Complex, coherent and highly organized fluctuations are often seen in the data. Power spectrum analysis is a useful technique for measuring the spatiotemporal properties of these fluctuations. This complements the superb ionospheric imaging capabilities of the MWA, which is the first radio telescope capable of imaging the ionosphere. Typical angular displacements at 154 and 183\,MHz are 10--20\,arcsec, sub-pixel most of the time but systematically larger than expected from position fitting errors. The MWA is built on the planned site of the low-frequency portion of the future Square Kilometre Array \citep[SKA-low;][]{Dewdney2009}. It therefore has an important role to play in establishing the types and levels of ionospheric activity that will be experienced by SKA-low, and the analysis methods described here can be applied to this end.

This work only investigated fluctuations in the angular position of point sources. However, flux density is another quantity of astronomical interest that may also be affected by the ionosphere. Ionospheric scintillation is expected to be induced by irregularities whose transverse spatial scales are several kilometers or less, which is below what we can directly measure using the technique described here. The large-scale fluctuations we do detect should only produce negligible changes in the flux density, much smaller than the pixel-to-pixel RMS at typical instrument sensitivities. The current analysis therefore does not probe the TEC fluctuations that may give rise to significant amplitude variability. We intend to address this more quantitatively in a forthcoming paper.

Our data show evidence for slow-moving, alternating fronts of TEC depletion and enhancement which are likely to be field-aligned density ducts, and also longer-wavelength, faster-moving waves which may be TIDs. A stationary, radially-inward distortion pattern detected in dataset B may arise from a combination of ionospheric and instrumental effects. The presence of coherent, time-dependent fluctuations in both datasets, in particular dataset B which was chosen for its relatively low amplitude of source angular displacements, implies that ionospheric effects contribute significantly to source motions in MWA data at 154 and 183\,MHz, and are present even in relatively undistorted datasets. This suggests that the successful implementation of algorithms targeting ionospheric distortions will likely lead to significant improvements in astrometric accuracy for point-source observations with the MWA and other low-frequency arrays. 

The wide instantaneous FoV and the high completeness of pierce points enable the MWA to map ionospheric structure over a $\sim$150\,km region at a spatial resolution of $\sim$3\,km (300\,km altitude), bridging the gap between global-scale GPS measurements and local-scale radar measurements. Contemporaneous GPS, radar, airglow and MWA observations would enable geophysical plasma phenomena to be studied holistically on a wide range of length scales. Although airglow and MWA measurements probe similar spatial scales, the MWA is sensitive to a much larger range of vertical scales than airglow. In future, we intend to perform more extensive studies involving a larger number of MWA datasets, both to quantify the unwanted effects of the ionosphere on the primary data products for radio astronomy, and also to harness the valuable imaging capabilities of the MWA for the advancement of ionospheric science. 


%
%
%
%
%
%
%

\begin{acknowledgments}
The data supporting this paper are available upon request submitted via email to the corresponding author at sloi5113@uni.sydney.edu.au. This scientific work makes use of the Murchison Radio-astronomy Observatory. We acknowledge the Wajarri Yamatji people as the traditional owners of the Observatory site. Support for the MWA comes from the U.S. National Science Foundation (grants AST-0457585, PHY-0835713, CAREER-0847753, and AST-0908884), the Australian Research Council (LIEF grants LE0775621 and LE0882938), the U.S. Air Force Office of Scientic Research (grant FA9550-0510247), and the Centre for All-sky Astrophysics (an Australian Research Council Centre of Excellence funded by grant CE110001020). Support is also provided by the Smithsonian Astrophysical Observatory, the MIT School of Science, the Raman Research Institute, the Australian National University, and the Victoria University of Wellington (via grant MED-E1799 from the New Zealand Ministry of Economic Development and an IBM Shared University Research Grant). The Australian Federal government provides additional support via the National Collaborative Research Infrastructure Strategy, Education Investment Fund, and the Australia India Strategic Research Fund, and Astronomy Australia Limited, under contract to Curtin University. We acknowledge the iVEC Petabyte Data Store, the Initiative in Innovative Computing and the CUDA Center for Excellence sponsored by NVIDIA at Harvard University, and the International Centre for Radio Astronomy Research (ICRAR), a Joint Venture of Curtin University and The University of Western Australia, funded by the Western Australian State government. IHC acknowledges the support of grants DP110101587 and DP140103933. CMT is supported by an Australian Research Council DECRA Award, DE140100316. DLK acknowledges the support of grant AST-1412421.
\end{acknowledgments}

\appendix
\renewcommand{\thefigure}{A\arabic{figure}}
\setcounter{figure}{0}

\section{Simulated Data}\label{sec:pseudo}
We performed experiments with simulated data to assess the severity of sampling-related artifacts related to typical pierce-point patterns of the MWA. We tested two types of input: (1) a travelling plane wave, and (2) a spectrum of waves with amplitudes following a power law. To conduct the tests, we used the sampling pattern from dataset B, trimmed to dimensions of $200 \times 200 \times 60$ and with temporal gaps removed. 

\subsection{Experiment 1: Plane wave} \label{sec:pseudo1}
\begin{figure}
  \centering
  \includegraphics[width=0.49\textwidth]{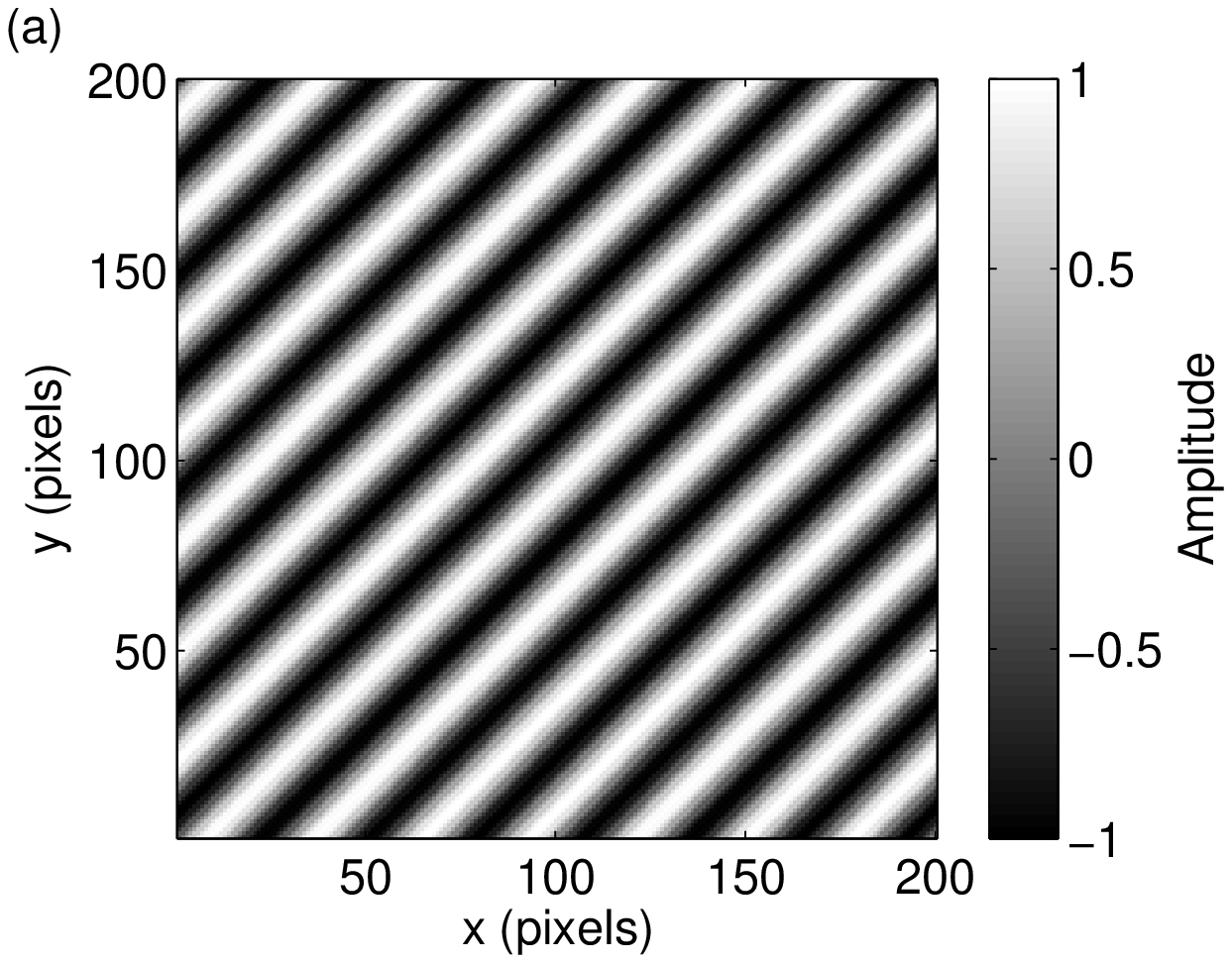}
  \includegraphics[width=0.49\textwidth]{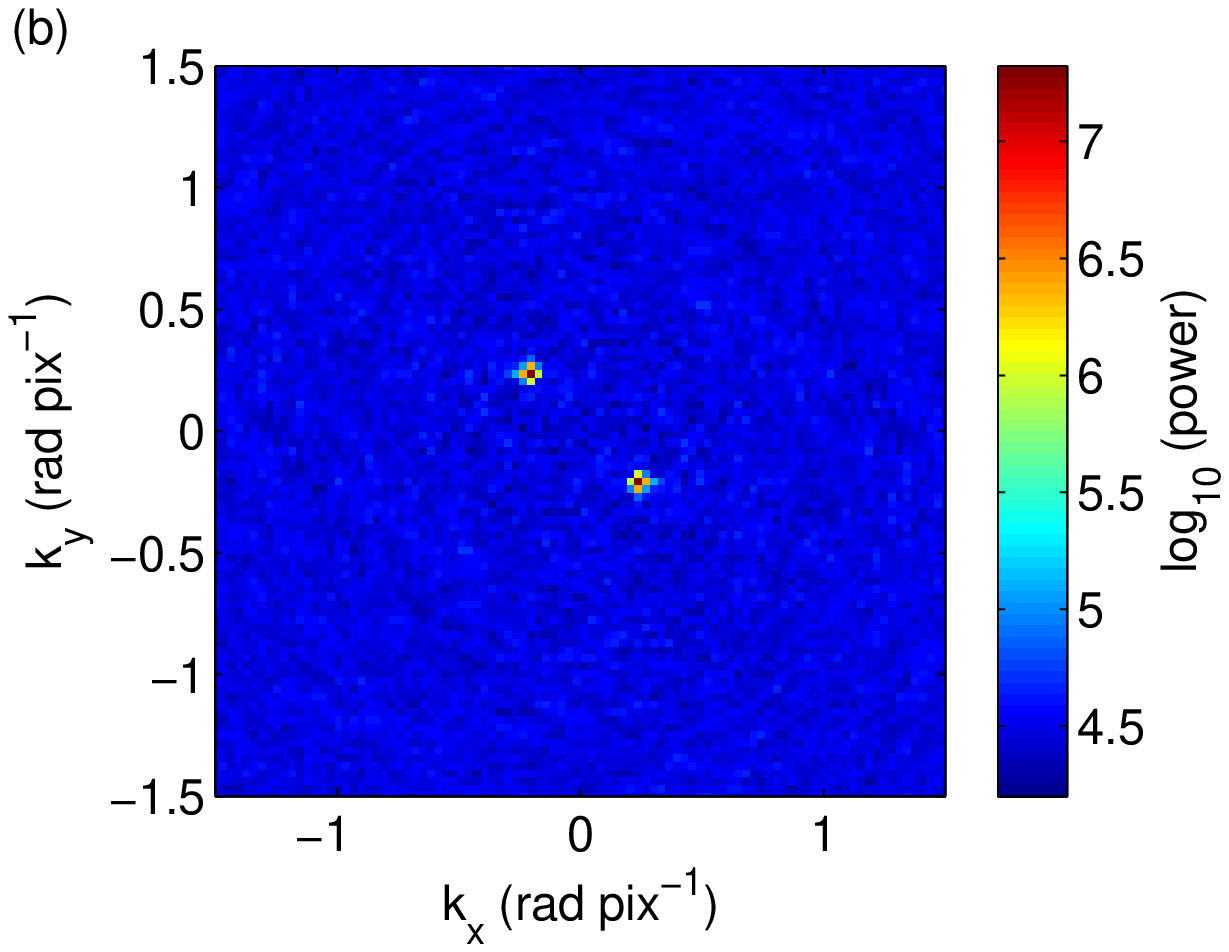}
  \includegraphics[width=0.49\textwidth]{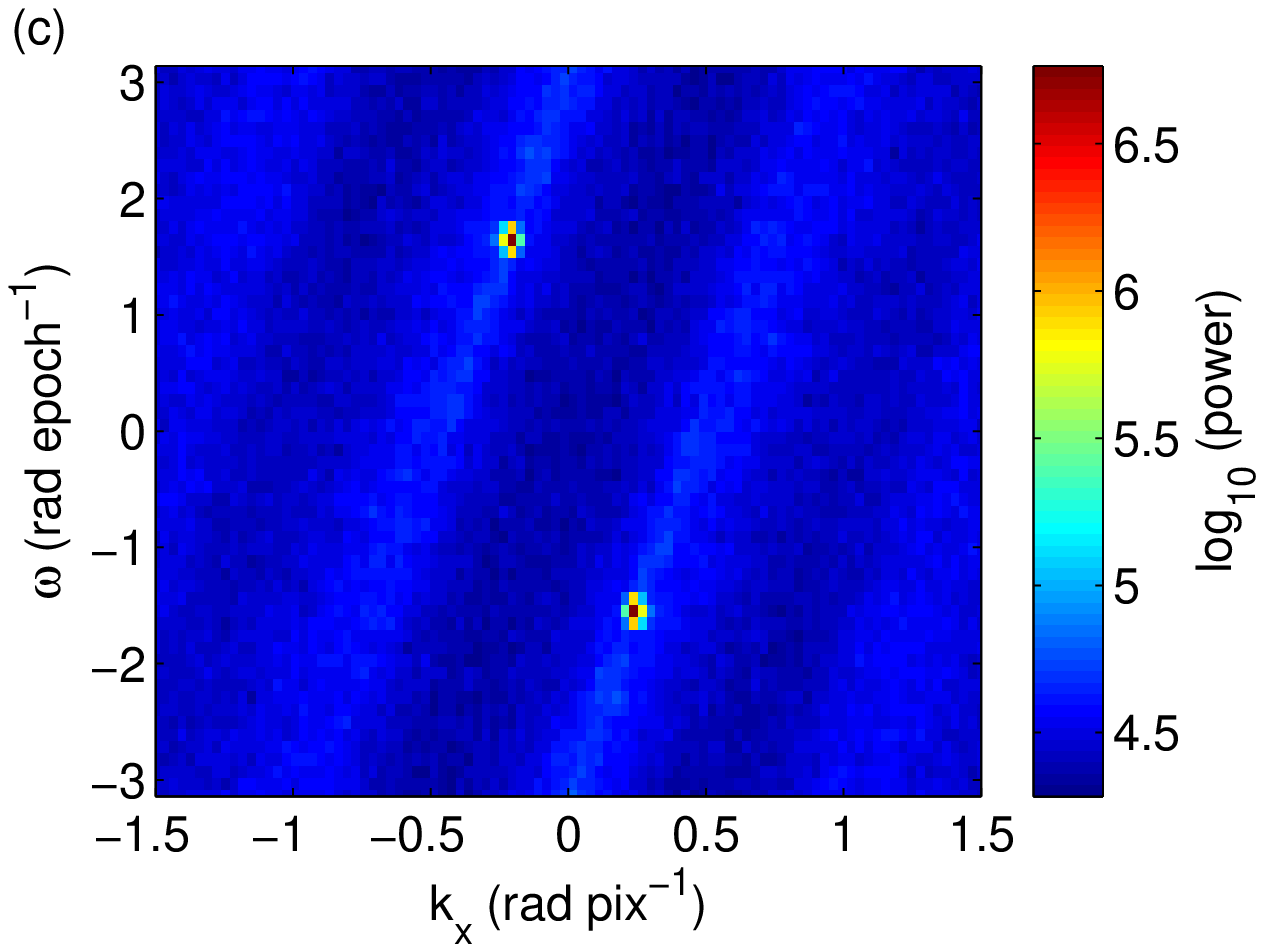}
  \caption{(a) Snapshot of input plane wave for experiment 1, which has a wavelength of 20 pixels and a period of 4 epochs, and the output power spectrum averaged over (b) $\omega$ and (c) $k_y$ showing a peak at $|k| = 0.32$ rad pix$^{-1}$, $\omega = 1.6$ rad epoch$^{-1}$. The diagonal ripples in (c) are features of the response function for dataset B, whose sampling distribution was borrowed for this experiment.}
  \label{fig:planewave}
\end{figure}

To test the system response we injected a single plane wave with a wavelength of 20 pixels and period of 4 epochs, travelling diagonally towards the bottom-right in Figure \ref{fig:planewave}a. The resulting power spectrum is shown in Figs \ref{fig:planewave}b and \ref{fig:planewave}c, where we have collapsed the datacube down to two dimensions for visualization by averaging over either $\omega$ or $k_y$. 

There are strong peaks at $|k| \approx 0.32$ rad pix$^{-1}$ and $\omega \approx 1.6$ rad epoch$^{-1}$, implying a wavelength of $\sim$20 pixels and a period of $\sim$3.9 epochs, in close agreement with the input values. Note that the power spectra are inversion-symmetric and so a given wave mode produces a pair of peaks on opposite sides of the origin, corresponding to positive and negative frequencies of equal amplitude. The direction of propagation can also be recovered from the power spectra: the line joining the peaks in Figure \ref{fig:planewave}b has $\mathrm{d}k_y/\mathrm{d}k_x = -1$, which matches the input propagation angle of $-45^\circ$, modulo 180$^\circ$ (Figure \ref{fig:planewave}a). The wave is therefore either moving towards the top left or the bottom right. This can be narrowed down to a single direction by considering the line joining the peaks in Figure \ref{fig:planewave}c, which has $\mathrm{d} \omega/\mathrm{d} k_x < 0 \implies \mathrm{d} x / \mathrm{d} t = -\mathrm{d} \omega / \mathrm{d} k_x > 0$, i.e.~the phase velocity has a positive $x$-component. Hence we conclude that motion is towards the bottom-right.

\subsection{Experiment 2: Power law} \label{sec:pseudo2}
We tested a superposition of modes with amplitudes following a simple isotropic power law, where the power (squared modulus) of a certain mode with spatial and temporal frequencies of $k$ and $\omega$, respectively, is given by
\begin{equation}
  \mathcal{P}(\omega, k) \propto k^m \omega^n \:. \label{eq:powerlaw}
\end{equation}

\begin{figure}
  \centering
  \includegraphics[width=0.49\textwidth]{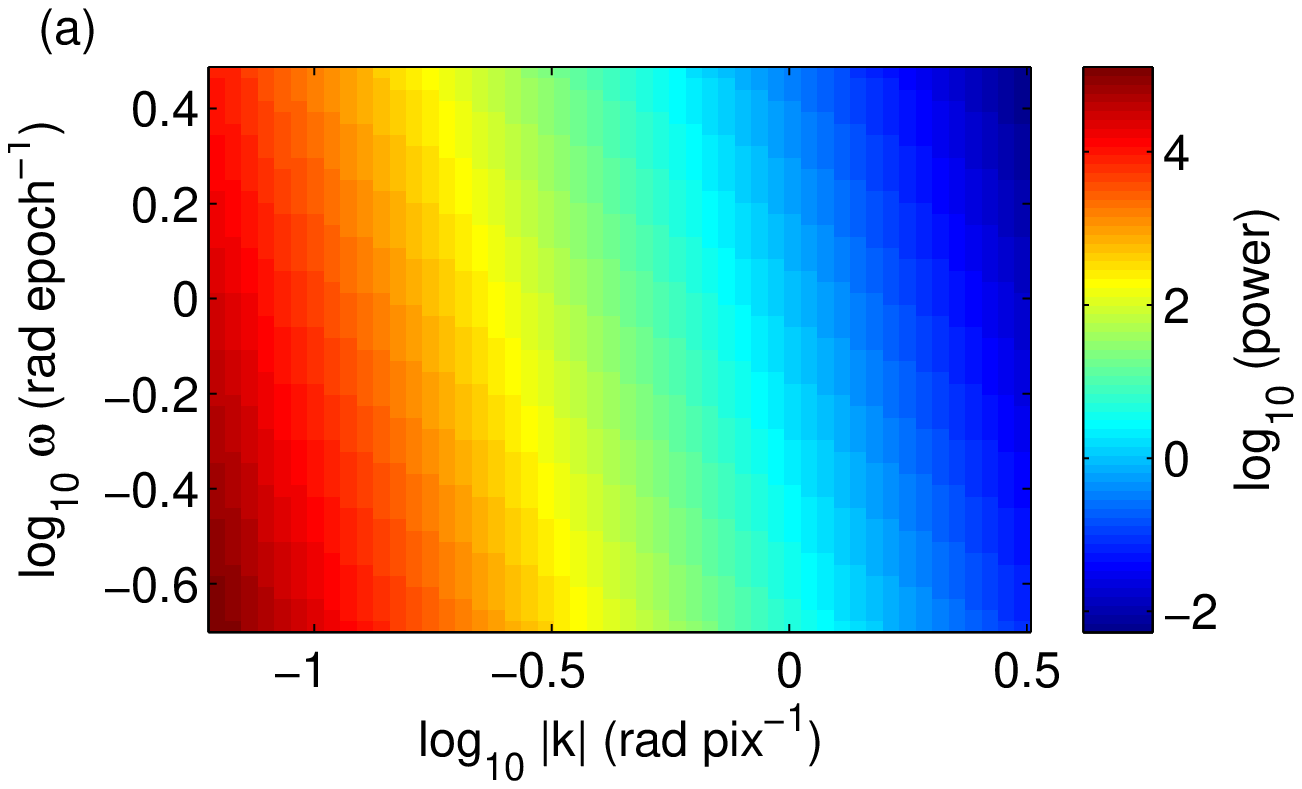}
  \includegraphics[width=0.49\textwidth]{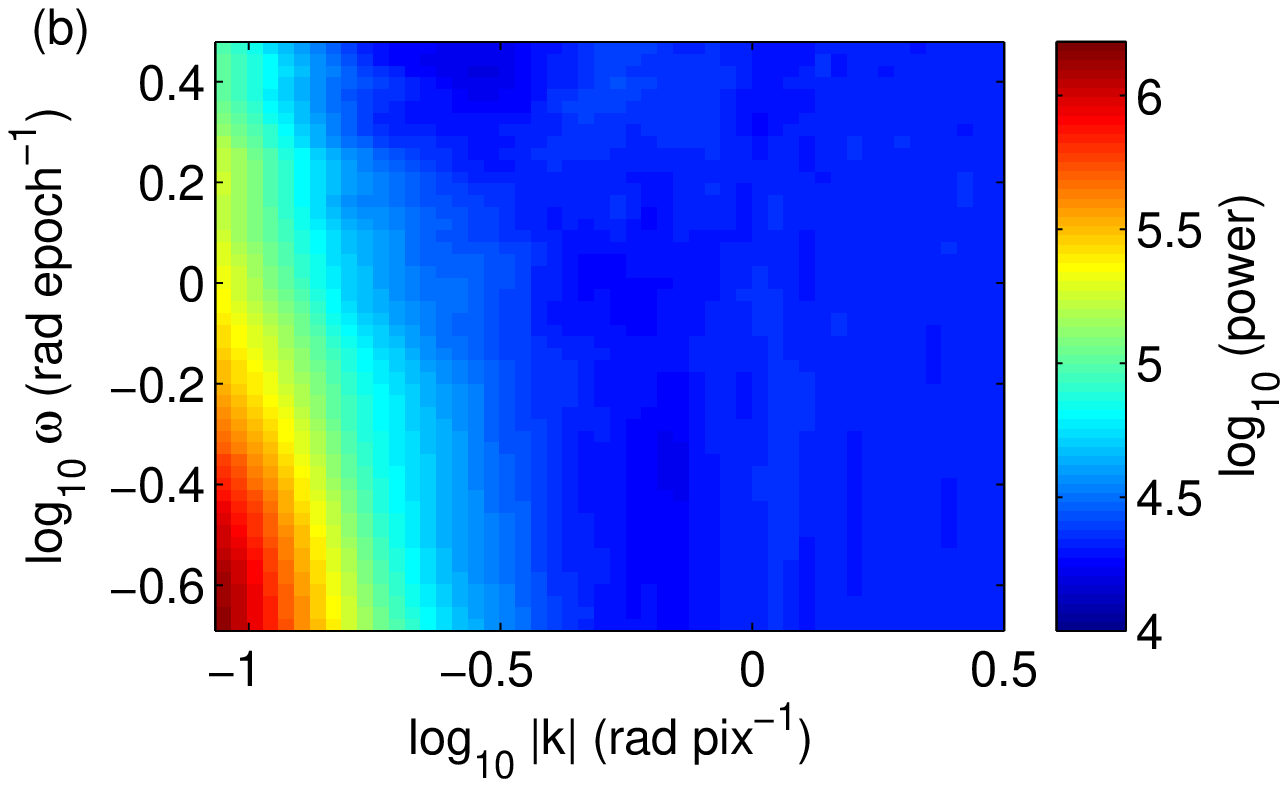}
  \caption{(a) The input power spectrum for experiment 2, and (b) the output power spectrum, shown on logarithmic axes. The output spectrum has been boxcar-smoothed with a $ 3\times 3$ pixel element with respect to the linear grid. Note the use of logarithmic axes, which are best for visualizing a power law.}
  \label{fig:powerlaw}
\end{figure}

The input and output power spectra for $m = -11/3$ and $n = -1$ are shown in Figure \ref{fig:powerlaw}, interpolated to a logarithmic grid. The flatness of the output spectrum compared to the input is likely to be a result of spectral leakage degrading the S/N ratio. However, the qualitative features of the input are recovered reasonably well.

\end{article}

\begin{table}[H]
  \centering
  \caption{Parameters for the two MWA datasets.}
    \begin{tabular}{ccc} \hline
    Parameter & Dataset A & Dataset B \\ \hline
    Central frequency (MHz) & 183 & 154 \\
    Bandwidth (MHz) & 30.72 & 30.72 \\
    Polarization & Stokes I & Stokes I \\
    Calibrator & 3C444 & PKS 2356$-$61 \\
    No.~of snapshots & 38 & 77 \\
    Missing data & None & Snapshots 57 and 61 \\
    Snapshot cadence (min) & 2 & 6 \\
    Integration time per snapshot (s) & 112 & 112 \\
    Span of observations (hr) & 1.3 & 8 \\
    UTC start & 2014-02-05 15:13:51 & 2013-09-16 13:30:39 \\
    UTC end & 2014-02-05 16:29:03 & 2013-09-16 21:24:39 \\
    AWST\tablenotemark{a} start & 2014-02-05 23:13:51 & 2013-09-16 21:30:40 \\
    AWST end & 2014-02-06 00:29:03 & 2013-09-17 05:24:40 \\
    LST start & 08:03:18 & 21:00:00 \\
    LST end & 09:18:43 & 04:55:19 \\
    Pointing center (J2000) & RA, Dec = $150^\circ, -24\fdg8$ & Dec = $-55\fdg0$ along meridian \\
    Pixel resolution (arcsec) & 45 & 45 \\
    Synthesized beam FWHM (arcsec) & 119 & 130 \\
    Visibility time resolution (s) & 4 & 1 \\
    Visibility frequency resolution (kHz) & 40 & 160 \\ \hline
    \end{tabular}
    \tablenotetext{a}{Australian Western Standard Time (UTC +8)}
  \label{tab:datasets}
\end{table}

\end{document}